\documentclass{statsoc}

\usepackage[a4paper]{geometry}
\usepackage{etoolbox}
\usepackage{float}

\makeatletter
\patchcmd{\@makecaption}
  {\parbox}
  {\advance\@tempdima-\fontdimen2} 
  {}{}
\makeatother

\usepackage{graphicx}       
\usepackage{amsthm,amsmath,amsfonts,amssymb} 

\usepackage{makecell}
\usepackage{arydshln}

\def \N{\mbox{N}}

\newcommand{\bp}{\mbox{\boldmath $p$}}

\newcommand{\bolds}{\mbox{\boldmath $s$}}

\newcommand{\bx}{\mbox{\boldmath $x$}}

\newcommand{\bms}{\mbox{\boldmath $s$}}
\newcommand{\bmt}{\mbox{\boldmath $t$}}
\newcommand{\bmy}{\mbox{\boldmath $y$}}
\newcommand{\bmx}{\mbox{\boldmath $x$}}
\newcommand{\bmX}{\mbox{\boldmath $X$}}

\newcommand{\bmY}{\mbox{\boldmath $Y$}}
\newcommand{\bmT}{\mbox{\boldmath $T$}}
\newcommand{\bmM}{\mbox{\boldmath $M$}}

\newcommand{\bmepsilon}{\mbox{\boldmath $\epsilon$}}
\newcommand{\bmuu}{\mbox{\boldmath $u$}}
\newcommand{\bmmu}{\mbox{\boldmath $\mu$}}
\newcommand{\bmtt}{\mbox{\boldmath $\theta$}}
\newcommand{\bbeta}{\mbox{\boldmath $\beta$}}

\usepackage{tikz}
\usetikzlibrary{positioning}
\usetikzlibrary{calc}
\usetikzlibrary{shapes,fit} 
\usetikzlibrary{arrows,decorations.markings} 
\usepackage{scalerel}

\title[Modeling U/R Fractions]{Modeling Urban/Rural Fractions in Low- and Middle-Income Countries}

\author[Yunhan Wu {\it et al.}]{Yunhan Wu}
\address{Department of Biostatistics, University of Washington, USA}
\author[Wu \& Wakefield]{Jon Wakefield}
\address{Department of Statistics and Department of Biostatistics, University of Washington, USA
}

\begin{document}

\begin{abstract}
In low- and middle-income countries, household surveys are the most reliable data source to examine health and demographic indicators at the subnational level, an exercise in small area estimation. Model-based unit-level models are favored in producing the subnational estimates at fine scale, such as the admin-2 level. Typically, the surveys employ stratified two-stage cluster sampling with strata consisting of an urban/rural designation crossed with administrative regions. To avoid bias and increase predictive precision, the stratification should be acknowledged in the analysis. To move from the cluster to the area requires an aggregation step in which the prevalence surface is averaged with respect to population density. This requires estimating a partition of the study area into its urban and rural components, and to do this we experiment with a variety of classification algorithms, including logistic regression, Bayesian additive regression trees and gradient boosted trees. Pixel-level covariate surfaces are used to improve prediction.  We estimate spatial HIV prevalence in women of age 15-49 in Malawi using the stratification/aggregation method we propose.
\end{abstract}

\section{Introduction}

Most low- and middle-income countries (LMIC) are not covered by Civil Registration and Vital Statistics (CRVS) systems. Consequently, the estimation of health and population indicators relies on
other types of data, in particular household surveys such as Demographic Health Surveys (DHS), Malaria
Indicator Surveys (MIS) and Multiple Indicator Cluster Surveys (MICS). These surveys typically support reliable estimation for key survey indicators at the national level and the first subnational administrative level (states, provinces, or regions), namely admin-1 areas. 

While estimates at national level and admin-1 level serve as important measures for policy makers, they do not reflect subnational estimation at finer scale, such as at admin-2 level, where health programs are designed and implemented \citep{mayala:etal:19}. In recent years, there has been a growing need for more localized DHS estimates. Formal use of local data can help in determining district
health priorities and planning, allocating resource, 
managing health workers to better serve the needs of the local population \citep{wickremasinghe:etal:16}.
Beyond local needs, international development goals also boost the
demand for admin-2 estimates \citep{utazi:etal:21}. Sustainable Development Goals (SDGs) requires an improvement in the measurement and understanding of local geographic
patterns to support more region-customized decision-making and more efficient program implementation
\citep{united:nations:general:assembly:15}. Reliable estimates at admin-2 level facilitate the effort to improve health outcomes for all and reduce within-country inequalities, which are a priority of the SDGs \citep{hosseinpoor:etal:15}.

Producing area-level estimates for key indicators is within the sphere of small-area estimation (SAE). 
There are two categories of models in the SAE literature: area-level models and unit-level (or cluster-level) models. Area-level models begin with design-based estimates. The DHS and MICS surveys use a stratified, two-stage cluster design with
probability proportional to size (PPS) sampling. Area-level models build upon weighted estimates with the complex designs being explicitly accounted for. In their seminal work, Fay and Herriot \citep{fay:herriot:79} propose a hierarchical models that incorporates random effects to improve the precision of weighted estimates. No additional considerations about the design are needed for area-level model because they are already incorporated in the weighted estimators, along with their design based variance estimates.  

\begin{figure}
    \centering
    \makebox{
    \includegraphics[width=0.95\linewidth,trim={0 3.2cm 0,  3cm},clip]{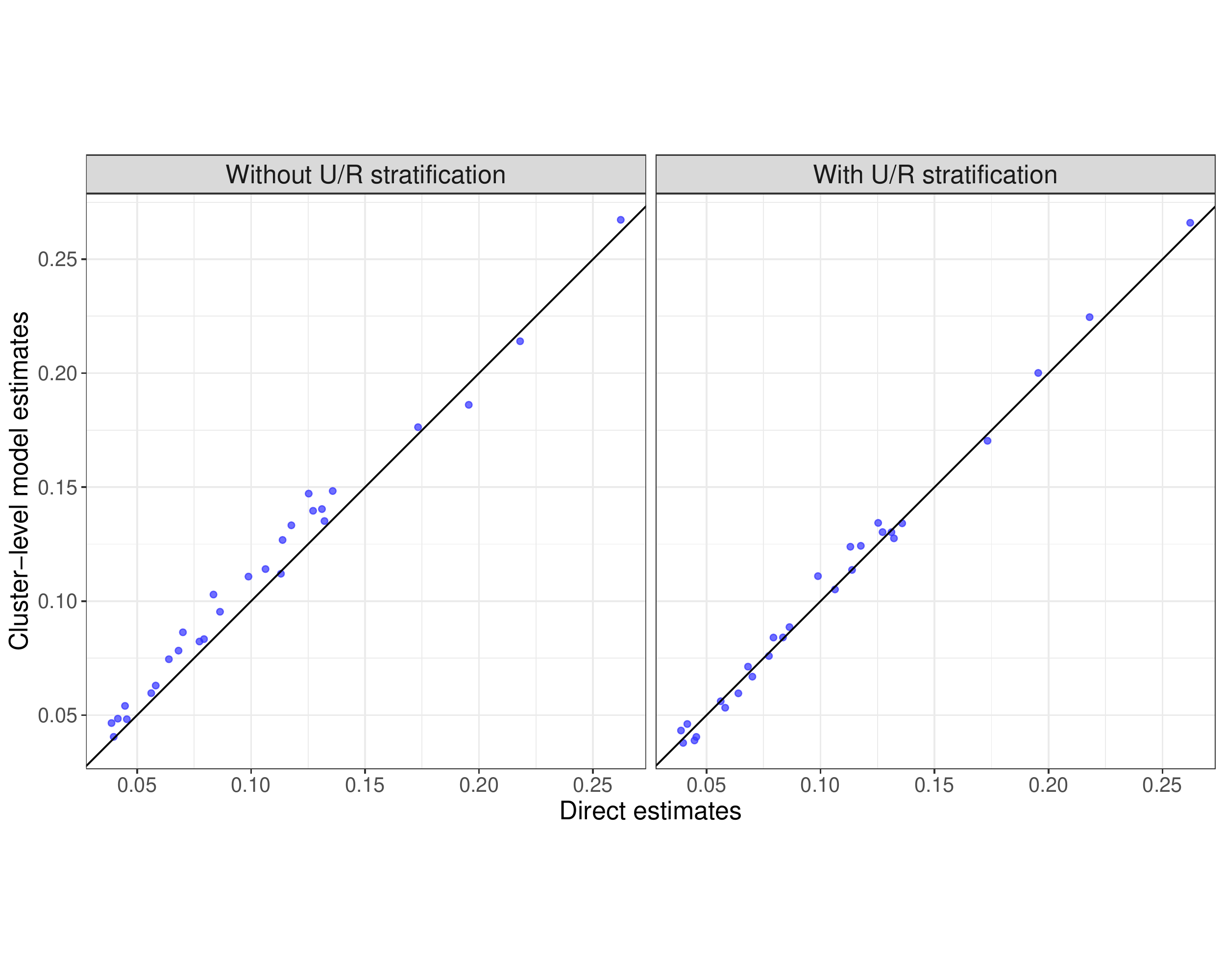}
    }
    \caption{Comparison of cluster-level model estimates with direct estimates for HIV prevalence amongst women of age 15-49 in admin-2 regions of Malawi. Estimates from left/right panels are from fixed effects cluster level model without/with urban/rural stratification.}
        \label{fig:stratified-compare-2}
\end{figure}

The design based area-level models perform well in producing admin-1 estimates, as DHS is powered to admin-1 level. For example, \cite{li:etal:19} estimate under-5 mortality in 35 African countries using a space-time Fay-Herriot model. However, due to data sparsity, weighted estimates at admin-2 level are unstable, or unavailable. Such a scenario calls for a unit-level model-based approach for reliable estimates. \cite{battese:etal:88} introduce a nested error regression model, which models the outcome at the level of the sampling unit (cluster level in DHS surveys) and includes area-level random effects to reduce uncertainty in estimates. A major challenge for these models is that they do not explicitly acknowledge the design for the survey and failing to account for the stratification might incur bias \citep{dong:wakefield:21, paige2022design}. Notably,  two large-scale producers of subnational demographic and health estimates, IHME and WorldPop do not adjust for the U/R stratification. Unfortunately, accounting for the stratification is not straightforward, as we shall see.

In most DHS surveys, urban/rural (U/R) serves as a binary stratification variable. Often urban clusters are over-sampled, compared with rural clusters. This is particularly the case in less urbanized areas, because exact PPS sampling will result in insufficient samples for urban strata in a LMIC setting \citep{dhsmanual}. Moreover, the prevalence of outcomes of interest often differs in urban and rural areas. Consequently, ignoring the stratification might result in bias because of the non-representative nature of the sample. The extent of the bias is closely related to the extent of urban over-sampling and the association between the response and U/R status. We use the HIV prevalence of women of age 15-49 in Malawi in 2015-2016 to illustrate the problem. Each point in Figure \ref{fig:stratified-compare-2} represents the estimates from one admin-2 area. The Malawi DHS 2015-2016 is one of the few DHS surveys that is powered at admin-2 level, which makes it ideal to demonstrate the issue. As shown in the left panel, a fixed effects model (at admin-2 level) without U/R stratification produce estimates systematically higher than the survey weighted estimates. It is no surprise that such model overestimate the HIV prevalence because of the oversampling of urban samples in most admin-2 regions in Malawi and the fact that urban areas tend to have higher HIV prevalence than rural areas.

There are many important health indicators that exhibit urban and rural association. The list includes,  but is not limited to,  HIV prevalence, child stunting rate, child mortality rates and fertility rates. \cite{dong:wakefield:21} point out that appropriate adjustment for design in unit-level model requires: model terms that explicitly acknowledge the design, such as U/R fixed effects in the prevalence model; and an aggregation procedure that assembles strata specific estimates to area level estimates. The major challenge remains in obtaining the population fractions used for aggregation. A further complication is that the reported cluster locations are jittered version of the true locations - further details of this aspect are addressed in Section \ref{sec:jitter}. In this paper, we formalize a pipeline that account for stratification in unit-level models based on household surveys.  

The structure of the paper is as follows. In Section \ref{sec:modeling}, we introduce the methodological framework with an emphasis on the population fractions used for aggregation. In Section \ref{sec:results}, we presents results for both the classification model and the overall modeling procedure in the context of HIV prevalence in a space-only setting. We summarize the paper with a discussion in Section \ref{sec:discussion}. In the Appendix, we implement an under-5 mortality rate example to demonstrate how the method works for a space-time indicator. We also place the additional modeling choices and minor details in the Appendix.


\section{Modeling}\label{sec:modeling}

To produce prevalence estimates at the admin-2 level, we use an {\it aggregation model} that combines a {\it prevalence model} at the cluster level with an {\it U/R classification model}. The prevalence model and the classification model can be developed separately and the relationship between the components is shown in Figure \ref{fig:model}.

\tikzstyle{decision} = [diamond, draw, fill=blue!20, 
    text width=5.5em, text badly centered, node distance=3cm, inner sep=0pt]
\tikzstyle{block} = [rectangle, draw, fill=blue!20, 
    text width=10em, text centered, rounded corners, minimum height=4em]
\tikzstyle{line} = [draw, -latex']
\tikzstyle{cloud} = [draw, ellipse,fill=red!20, node distance=3cm, text width=4.5em,text centered,
    minimum height=2em]
    
 \begin{figure}

     \begin{center}
    \begin{small}
    
\begin{tikzpicture}[node distance = -.3cm, auto]
    \node [block] (level1left) {Prevalence\\ Model};
        \node [block, xshift=5.5cm] (level1right) {U/R\\ Classification Model};
    \node [block, below of=level1left, yshift=-3cm, xshift=2.75cm] (level2) {Aggregation\\ Model};

    \path [line] (level1left) -- (level2);
        \path [line] (level1right) -- (level2);
   
\end{tikzpicture}

    \caption{Overview of modeling process}

\label{fig:model}

\end{small}
     \end{center}
\end{figure}
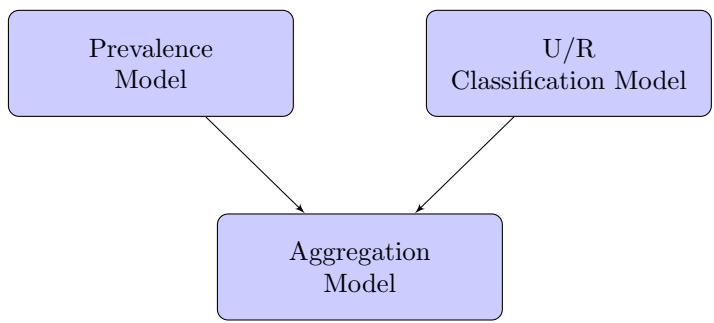

The U/R status of any point in the country is defined through the sampling frame, as we will further explain in Section \ref{A-B-surface}. Thus, the U/R classification is independent of the survey indicators. At the final stage, the outputs from the two models can be combined to form the aggregated estimates. 

We start by describing various prevalence models in Section \ref{prev-model} and provide details of the aggregation method in Section \ref{aggre-model}. Section \ref{sec:UR-class-frac} features the emphasis of the paper -- the U/R classification model and \ref{sec:jitter} considers the impact of jittering.

\subsection{Prevalence Model}
\label{prev-model}

We focus on unit-level models for prevalence estimation - these are generally required for admin-2 estimates in LMICs. The proposed framework aims to account for the design in particular with respect to the U/R stratification. In the setting of DHS surveys from LMICs, the units refer to clusters. We adopt Bayesian hierarchical models and use smoothing to produce estimates when the data are sparse. We start with a space only model for binary survey indicators and extend the model to a space-time setting. The association with the U/R stratification is incorporated as a fixed effect. In addition, we also consider including separate spatial fields for U/R, thereby allowing U/R effects to vary in subnational regions.

Here we adopt discrete spatial fields at the area level, though give a discussion about continuous spatial models in the Appendix. Moreover, we do not include covariates in the prevalence model here, but we do consider this extension in the Appendix.

We describe spatial prevalence model that we will use for the application of HIV prevalence for women of age 15-49 in Malawi.  We let $Z_c$ and $n_c$ represent the number of events and the number sampled in cluster $c$, which has location $\bolds_c$ (which we assume for the moment is the true location).
In the mixed effects model we assume,
\begin{eqnarray}\label{eq:binomial}
Z_{c} | p_{c} &\sim& \mbox{Binomial}(n_{c},p_{c})
\nonumber \\
p_c  &=& \mbox{expit} \left(  \alpha^{\text{rural}}\times I( \bms_c \in \mbox{rural}) + \alpha^{\text{urban}}\times I( \bms_c \in \mbox{urban}) + S(\bolds_c)
+\epsilon_c \right) \label{eq:fixed-effect-prev}
\end{eqnarray}
where $ \alpha^{\text{rural}}$ and $\alpha^{\text{urban}}$ are intercepts for rural and urban with an associated indicator to determine whether $\bolds_c$ is in a part of area $i$ designated as rural or urban, $S(\bolds_c)$ is a spatial random effect associated with location $\bolds_c$, and $\epsilon_c \sim_{iid} \N(0,\sigma_\epsilon^2)$ is a cluster-specific independent random effect. We assume the latter is reflecting the overdispersion (excess-binomial variation) due to within-cluster correlation; for more discussion, see \cite{dong:wakefield:21}.

We model the spatial random effect term in equation (\ref{eq:fixed-effect-prev}) as the summation of an independent and identically distributed distributed component and a spatial component, such that $S(\bolds_{c})=e_{i[\boldsymbol{s}_c]}+S_{i[\boldsymbol{s}_c]}=b_{i[\boldsymbol{s}_c]}$. The notation $i[\boldsymbol{s}_c]$ means the area within which the cluster $\bolds_{c}$ resides. We model the iid random effect as $e_i \sim_{iid} \N(0,\sigma^2_e)$, the spatial part as an intrinsic conditional autoregressive (ICAR) component and $\boldsymbol{b}=[b_1,\dots,b_{m}]$ so that $ \boldsymbol{b} \sim \mbox{BYM2}(\sigma_s^2,\phi)$. We adopt the BYM2 parameterization \citep{riebler:etal:16} for the spatial random effects, in which $\sigma_s^2$ represents the marginal variance and $\phi$ represents the proportion of this variance explained by the spatial component. We impose penalized complexity (PC) priors \citep{simpson:etal:17} on the variance and correlation parameters $\sigma_s^2$ and $\phi$. In a simplified setting where we drop the spatial component of the random effects, the prevalence model will reduce to an iid random effect model.

Under the model specified in equation (\ref{eq:fixed-effect-prev}), we obtain the strata (area $\times$ U/R) specific risk in area $i$ as,
\begin{eqnarray}
p_{i,R} &=& \mbox{expit} \left( \alpha^{\text{rural}}+ b_i \right)
\nonumber \\
p_{i,U} &=& \mbox{expit} \left( \alpha^{\text{urban}}  +b_i \right),
\label{strata-risk}
\end{eqnarray}
for rural and urban prevalence in area $i$.
This calculation is straightforward since we have assumed a constant risk within the area.

In a more flexible model, we allow distinct spatial fields for urban and rural stratum, where the U/R effect is allowed to vary across admin areas,
\begin{eqnarray}\label{eq:binomial}
Z_{c} | p_{c} &\sim& \mbox{Binomial}(n_{c},p_{c})
\nonumber \\
p_c  &=& \mbox{expit} \left[ (\alpha^{\text{rural}}+  b_i^\text{rural})\times I( \bms_c \in \mbox{rural}) + \right. 
\nonumber \\
&&\left. (\alpha^{\text{urban}}+  b_i^\text{urban})\times I( \bms_c \in \mbox{urban})+\epsilon_c 
 \right]
\label{eq:inter-prev}
\end{eqnarray}
The corresponding strata specific risk is,
\begin{eqnarray}
p_{i,R} &=& \mbox{expit} \Bigl( \alpha^{\text{rural}} + b_i^\text{rural}\Bigr)\nonumber \\ 
p_{i,U} &=& \mbox{expit} \Bigl( \alpha^{\text{urban}}+ b_i^\text{urban}\Bigr)
\label{int-strata-risk}
\end{eqnarray}

\subsection{Aggregation Model}
\label{aggre-model}

For a generic binary indicator $Z$, to calculate the area-level prevalence in area $i$, $p_i$, we aggregate the urban and rural specific risks:
\begin{eqnarray}
p_i &=&\underbrace{ \Pr( Z=1 | \mbox{rural},i  )}_\text{Prevalence Model}
\times\underbrace{ \Pr(\mbox{rural}|i) }_\text{Classification Model}
+ \underbrace{ \Pr( Z=1 | \mbox{urban},i)}_\text{Prevalence Model}\times \underbrace{\Pr(\mbox{urban}|i)}_\text{Classification Model}
\nonumber \\
&=& p_{i,U}\times q_{i} + p_{i,R}\times (1-q_{i})
\end{eqnarray}
where $q_i$ is the urban fraction of the subpopulation corresponding to $Z$, in area $i$. We will now discuss in detail the estimation of $q_i$.
\subsection{U/R Classification and Fraction Model}
\label{sec:UR-class-frac}
\subsubsection{Overview}
\label{A-B-surface}

In a model-based approach, we need to adjust for urban and rural when the outcome depends on U/R and the survey did unequal sampling with respect to urban and rural. We emphasize the distinction between DHS U/R stratification and the ``true'' urban and rural classification over time. The definition and allocation of urban and rural clusters in a DHS survey occurs at the time of the census, which is the usual source of the {\it sampling frame}. The urban and rural cluster map is constructed whenever the sampling frame is formed, and is constant over time,  just as the weights are in a design-based framework. However, the populations in those areas do change over time. 


In a perfect world, we would have access to all clusters in the sampling frame, along with the relevant populations in these clusters. Based on the U/R status of those clusters, we would be able to calculate U/R fractions. However, such information is rarely available. Instead, we propose a generic pipeline using information from the sampled cluster and the Worldpop population density surface \citep{stevens2015disaggregating}. The main idea is to use pixels to represent clusters in the sampling frame. We construct a pixelated U/R map to estimate the actual U/R surface in the sampling frame; and use the Worldpop population density to allocate the relevant population in the pixels. We construct the U/R map using a classification model, which we will discuss in Section \ref{sec:classification}. For the moment assume we have a classification map $C$ and a generic population density map $H$ for the study region. Let $g$ be the index of pixels for a gridded surface and $\bms(g)$ be the spatial coordinates of pixel $g$. The classification and population density at pixel $g$ are $C_{\scaleto{\bms}{4pt}(g)}\equiv C_g$ and $h_{\scaleto{\bms}{4pt}(g)}\equiv H_g$ respectively. 
Then we can calculate the urban fraction in a generic area as, 
\begin{equation}
q = \frac{\sum I(C_g = \text{urban}) \times  H_{g}}{\sum H_{g}}
\label{eq:classify-to-frac}
\end{equation}

The method we propose is very flexible in that it allows us to calculate urban fraction for any sub-population and for any time. The estimation procedure is valid as long as we are working under the stratification defined by the specific sampling frame. Worldpop \citep{stevens2015disaggregating} provides yearly population density file for age-sex combinations in individual countries around the globe, and such data availability guarantees our methods are widely applicable.

\subsubsection{Classification Model}
\label{sec:classification}

The generic geographical problem is to classify each pixel on a grid over the study region as either urban or rural. Because different countries have distinct urban definitions, we normally define the study region as a single country, which coincides with the prevalence model. 

There exists a non-statistical method to classify urban pixels.  Motivated by the pattern that urban areas tend to be highly populated, \cite{wu:etal:21} adopt a thresholding algorithm to assign U/R labels for pixels merely based on population density. As DHS surveys often use a census as the sampling frame, it is likely that national and/or subnational urban total population fractions for the frame are available. We emphasize that those fractions are with respect to the total population, which is different from we aim to obtain: urban fractions for a specific age-sex subpopulation, such as women of age 15-49 in our HIV prevalence example. Moreover, the subnational urban fractions are usually only available at a level higher than we desire, e.g., admin-1. 

Combining the fractions with WorldPop population density surfaces \citep{stevens2015disaggregating} for total population from the year of the sampling frame, we can derive an area-specific population density threshold which yields the correct proportion at the admin-1 level \citep{wu:etal:21}. To be specific, if we know the urban fractions for area $A_i$ is $r_i$, then pixels in area $i$ with population density greater than $N_{[i]}$ are classified as urban (with the rest as rural), with $N_{[i]}$ determined through the following thresholding algorithm,  
\begin{equation}
    N_{[i]}=\underset{N_{[i]}}{\mathrm{argmin}} \left| \frac{\sum_{g \in A_i} I{ (N_{g}>N_{[i]} )} \times  N_{g}}{\sum_{g \in A_i} N_{g}}- r_i \right|,
\label{eq:benchmark:pop}
\end{equation}
where $g$ indexes pixels within admin region $A_i$, such that $ \bms_g \in A_i$ and $N_{g}$ is the total population density at pixel $g$ (obtained from Worldpop).

The above method, which we refer as `population thresholding', is computationally efficient, but less reliable at finer spatial resolution. It is insufficient to only account for population density because there are many more criteria for defining urban settlements \citep{UNYearBook2018} and using a purely deterministic algorithm lacks statistical justification.

In this paper, we propose a more sophisticated pipeline that incorporates more data to better resemble the true U/R definition and our method is grounded with a legitimate statistical framework. The training data we consider are cluster information from DHS surveys, which contain the reported geographical coordinates and U/R status of each cluster. Those pixels only cover a small portion of the all gridded cells. As we are constructing an U/R surface consistent with the sampling frame, clusters from multiple DHS surveys can all be included in the training data as long as they are based on the same sampling frame. For example, both DHS Malawi 2010 and DHS Malawi 2015-16 used as sampling frame the 2008 Malawi Population and Housing Census, so we can combine the clusters to form a larger training set. 

Our classification model is built on a suite of covariates. From a scientific perspective, we seek to find covariates that better mimic the official U/R definition for each country. According to Table 6 from the Demographic Year Book 2018 \citep{UNYearBook2018}, important urban features include, but are not limited to, high population density, infrastructure development (such as access to electricity) and existence of local government, engaged in non-agricultural activities. From a modelling perspective, the covariates need to be available for DHS countries, and in reasonable geographical resolution (for example, 1km$\times$1km). Based on the above information, we consider the list of covariates in Table \ref{table:cov_summary}. They are available in raster format and we also indicate the data source. For time-varying covariates, we pick the year of the sampling frame. In addition, we include admin region as a predictor because U/R classification might be different in different parts of the country. Usually we include admin-1 region because not all admin-2 regions have clusters for most surveys.  

It is tempting to adopt spatial models like stochastic partial differential equations (SPDE) for constructing a U/R surface. However, urban areas often have hard boundaries and modeling a binary surface using a continuous spatial model is not appealing. In addition, we hope that the rich set of covariates will suffice to properly capture the spatial correlation between pixels. Hence, we opt for non-spatial statistical learning models, including Bayesian logistic regression, Bayesian additive regression trees (BART) \citep{chipman2010bart} and gradient boosted trees \citep{friedman2001greedy}. Under the Bayesian framework, it is straightforward to quantify uncertainty in classification models.

\begin{table}
\caption{\label{table:cov_summary}Summary of covariates.}
\centering
\fbox{%
\begin{tabular}{|l|l|l|} 
\hline
 Covariate Name & Details & Source \\
 \hline
 population  & Population density & Worldpop \citep{stevens2015disaggregating} \\  \hdashline [0.5pt/5pt]
 night time light & \makecell[l]{Annual average visible, stable \\ lights and cloud free coverages} & NOAA \\ \hdashline [0.5pt/5pt]
 vegetation index & Mean annual terra vegetation indices & MOD13A3 v006 \citep{didan2015mod13a3}  \\ \hdashline [0.5pt/5pt]
 precipitation & Average annual precipitation & WorldClim \citep{fick2017worldclim} \\ \hdashline [0.5pt/5pt]
 elevation & Altitude & WorldClim \citep{fick2017worldclim} \\ \hdashline [0.5pt/5pt]
 temperature & Average annual temperature & WorldClim \citep{fick2017worldclim}  \\  \hdashline [0.5pt/5pt]
 access &  \makecell[l]{Travel time to the nearest city of \\ 50,000 or more people in 2000} & Product described in \cite{nelson2008travel}  \\ 
 \hline
\end{tabular}}
\end{table}


In the following, we use BART to illustrate the method. Details of other models can be found in the Appendix. Let $\bmY=\{Y_g\}$ denote the collection of U/R status with $Y=1/0$ indicating U/R for all pixels. Let $\bmx_g$ be the covariates at the pixel indexed by $g$. We use $\bmy_c$ and $\mathbf{X}_c$ to denote observed U/R status and covariates at DHS clusters, which constitute the training data. 

The sum-of-trees model from BART can be expressed as 
\begin{equation}
P(Y_g= 1|\bmx_{g},\bmtt) = \Phi [G(\bmx_{g},\bmtt)]
\end{equation}
where \begin{equation}
G(\bmx_{g},\bmtt)=\sum_{q=1}^{q_0}g(\bmx_{g};\bmT_q,\bmM_q)
\end{equation}
Here, we use the probit link binary data, following the original BART paper \citep{chipman2010bart}, the function $g$ represents a regression tree, $T$ denotes a tree structure and $M$ denotes the parameter values for each of the terminal nodes of $T$.

BART uses a backfitting algorithm to generating a sequence of draws of tree parameters $\bmtt=\{\bmT,\bmM\}=\{(\bmT_1,\bmM_1), ... ,(\bmT_{q0},\bmM_{q0})\}$ such that they will converge to the posterior of $p(\{\bmT,\bmM\}|\bmy,\{\bmx(\bms_c)\})=p(\bmtt|\bmy,\mathbf{X}_{\{\bms_c\}})$. Inference then can be made from $m_0$ posterior draws of tree parameters $\bmtt^{(m)}$, $m= 1, ..., m_0$.

Based on the classification model, we aim to determine the U/R status for all grid points within a country. The models we consider yield predicted probabilities and the quantities we aim to estimate are $P(Y_g= 1|\bmx_{g},\bmtt)$, where the predicted probability can be expressed as a function of the model parameters $\bmtt$. We may use a cutoff such as 0.5 to convert into binary classifications. Thus, we let $v_g(\bmtt)=P(Y_g= 1|\bmx_{g},\bmtt)$, where we drop the dependency on covariates because we can treat the covariates at pixel $g$ as known values. Under the Bayesian framework, we aim to estimate the posterior for $v_g(\bmtt)$, where $g=1,...,n_g$ and $n_g$ is the total number of pixels in the country. A point estimate can be derived from the posterior $\bmtt|\bmy$, such as the posterior mean, which we denote it as $\mu_g$,
\begin{align}
\begin{split}
E_{\scaleto{\bmtt|\bmy}{7pt}}[P(Y_g=1|\bmtt)] &=\int v_g(\bmtt) p(\bmtt|\bmy)d\bmtt \\
   & = \mu_g
\end{split}
\end{align}


We assign a classification for pixel $g$ at location $\bms_g$ as $I{ (\mu_{g}>\tau )}$, where $ \mu_{g}$ is the estimated probability of pixel $g$ being urban. Here the cutoff $\tau$ is a number in (0,1), and it may vary for different locations $g$, if additional information is available. 

Ideally, our classification surface would yield the same urban fraction(s) for total population as those from the sampling frame. Using a predefined and universal cutoff like $\tau=0.5$ will not in general produce predicted probabilities that are well calibrated. To address that, we adopt a strategy similar to the thresholding method described in equation (\ref{eq:benchmark:pop}). We refer to our deterministic calibration approach as exact benchmark. We consider the subnational case only and choose $\tau_i$ to make the estimated fraction closest to the subnational known proportions $r_i$, where $i$ indexes the admin regions.
For example, to match the admin-2 proportions in Malawi (these proportions are usually available only at admin-1 level), we have
\begin{equation}
    \tau_i=\underset{\tau_i}{\mathrm{argmin}} \left| \frac{\sum_{g \in A_i} I{ (\mu_{g}>\tau_i )} \times  N_{g}}{\sum_{g \in A_i} N_{g}}- r_i \right|
    \label{eq:exact-benchmark}
\end{equation}
where the summation is over all pixels $g$ within admin region $A_i$, such that $ g \in A_i$ and $N_{g}$ is the 2008 Malawi total population density at pixel $g$ (obtained from Worldpop).

The above derivation is based on the assumption that the posterior $\bmtt|\bmy$ is analytically available.
In practice, we use draws from the posterior to obtain the estimated predicted probability $\tilde{\mu}_g$ to $\mu_g$ as 
\begin{align}
\begin{split}
 \tilde{\mu}_g &=\frac{1}{m_0}\sum_{m=1}^{m_0} v_g(\bmtt^{(m)})
\end{split}
\label{eq:tilde_mu_g}
\end{align}
where $\bmtt^{(m)}$ are draws from the posterior $\bmtt|\bmy$ and $m_0$ is the number of posterior draws. Similarly, we estimate $\tau_i$ using $\tilde{\mu}_g$ such that 
\begin{equation}
\tilde{\tau}_i=\underset{\tau_i}{\mathrm{argmin}} \left| \frac{\sum_{g \in A_i} I{ (\tilde{\mu}_{g}>\tau_i )} \times  N_{g}}{\sum_{g \in A_i} N_{g}}- r_i \right|
\label{eq:tilde_tau}
\end{equation}
and in the end we classify the pixels as 
\begin{equation}
\tilde{C}_g = I(\tilde{\mu}_{g}>\tilde{\tau}_i) 
\label{eq:tilde_c_g}
\end{equation}
where $i$ is the index for the area such that $ \bms_g \in A_i$.

Finally, following equation (\ref{eq:classify-to-frac}), we may obtain estimates of urban fractions for population $N^{*}$ in an area $A_j^{*}$.
\begin{equation}
\tilde{q}_j = \frac{\sum_{g \in A^{*}_j} I(\tilde{C}_g = \text{urban}) \times N^{*}_g}{\sum_{g \in A^{*}_j} N^{*}_g}
\label{eq:classify-to-frac-est}
\end{equation}
where we adopt different notations ($N$ vs. $N^{*}$ and $A$ vs. $A^{*}$) to emphasize that the target population and domain are likely to be different from those in the classification model (we use total population as a covariate and for benchmarking). For example, we are able to obtain urban fraction estimates for women of age 15-49 in 2010 in admin-2 regions of Malawi.

\subsection{Jittering}
\label{sec:jitter}

To maintain the confidentiality of survey respondents, the DHS release a geographically displaced version of the true locations of clusters \citep{burgert2013geographic}. According to the rules, U/R clusters are displaced a distance up to 2/5 kms; with a further, randomly-selected 1\% of rural clusters displaced a distance up to 10 kms. The displaced distances are uniformly distributed and the displacement may not move the point across administrative boundaries (admin-1/2, country specific). Such displacement can harm our classification model. When a cluster is jittered to cross the U/R boundary, it becomes mislabeled. For example, an urban cluster in Blantyre from Malawi DHS 2015-2016 has been displaced into the mountains, as shown in Figure \ref{fig:cluster_529}(a). The true or undisplaced cluster location should be near the edge of the red circle, most likely in the northwest corner, because of high population density there. If we do not account for the displacement, the classification algorithm will be distorted. 


Following the development in \cite{wilson2020pointless}, we denote the relationship of the true and reported (jittered) cluster locations as 
\begin{equation}
\underbrace{\bmuu}_{\text{Reported Location}} = \underbrace{\bms}_{\text{True Location}} + \underbrace{\bmepsilon}_{\text{Displacement}}
\end{equation}
where $\bmepsilon$ is the result of the jittering probability density function. Under the DHS jittering algorithm, the true location is randomly displaced according to the distribution (in polar coordinates), \begin{equation}
p(\ell,\eta) = \frac{1}{2\pi R}\times I({0 <\ell <R})\times I({0<\eta<2\pi})
\end{equation}
where $R=2$km for urban clusters and $R=5$km for 99\% of rural clusters and $R=10$km for the remaining 1\% of rural clusters. As just discussed, the displaced location has to stay within the same strata, but we do not account for this in the notation, for simplicity. We describe two approaches to account for the jittering.

\subsubsection{Partial Adjustment}
\label{Partial-Adjustment}

\begin{figure}
    \centering
    \includegraphics[width=1\linewidth,trim={0 0cm 0cm 0},clip]{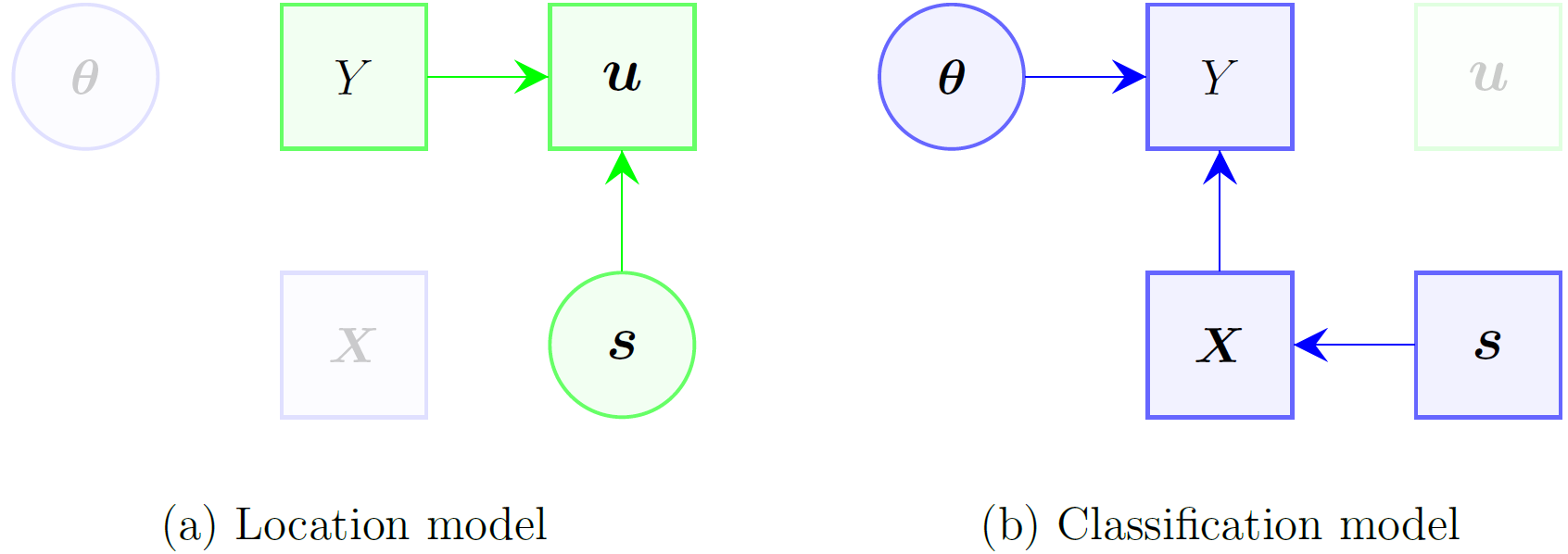}
      \caption{Graphical model for the partial adjustment method.}
      \label{fig:no-feed-graph}
\end{figure}

We first model the true locations of clusters without using information from the classification model, so that there is no feedback, as shown in Figure \ref{fig:no-feed-graph}. The true location of each generic cluster is estimated through density $p(\bms_c|\bmuu_c,y_c)$ and then we use the covariates  $\bmX_{\{\bms_c\}}$ based on the estimated locations to obtain estimates for $\bmtt$.

To statistically model the jittering process under a Bayesian setup, we impose a prior on a generic true location \citep{wilson2020pointless}. To ease notation, we drop index $c$ for cluster and specify the prior for a single cluster as 
\begin{align}
p(\bms = \bmt_k
 ) = d_k, \quad k=1,\dots,K, \label{eq:maskprior}
\end{align}
over potential locations $\{\bmt_k\}$, and 
where $d_k$ is the probability that potential location $\bmt_k$ was selected. Here, we assume the true cluster locations appear only on the center of gridded pixels, which is a discretization of the true sampling process. 
If probability proportional to size (PPS) sampling was undertaken (the usual strategy in the DHS), we may use $$d_k \propto N_k$$ where $N_k$ is the total population density of the cluster located at $\bmt_k$, obtained from the pixelated population density surface. In DHS surveys, the PPS sampling is often based on number of households and we use population density as a practical approximation. 
The posterior is then,
\begin{equation}
p(\bms = \bmt_k
|\bmuu, y
) \propto p(\bmuu | \bms = \bmt_k,y ) \times p(\bms = \bmt_k), \quad k=1,\dots,m,
\label{eq:true-loc-post}
\end{equation}
where $p(\bms)$ corresponds to (\ref{eq:maskprior}).

The distribution of $\bmuu | \bms$ is uniform on a circle of radius $R$, where $R$ depends on the U/R status $y$ of the cluster. We discretize the circle and use centers of pixels as potential locations for the cluster. We have
\begin{equation}
p(\bmuu | \bms = \bmt_k,R ) = C_{R} \times  I({0 < d(\bmuu, \bmt_k) < R})/{2\pi R d(\bmuu, \bmt_k)}
\end{equation}
where the distance between the pixelated candidate location
$\bmt_{k} = [t_{k1}, t_{k2}]$ and the reported location 
$\bmuu = [u_{1}, u_{2}]$ is
\begin{equation}
d(\bmuu, \bmt_k )= \left[(u_1 - t_{k1} )^2 + (u_2 - t_{k2} )^2\right]^{1/2}.
\end{equation}
The normalizing constant that accounts for the jittered point being restricted to stay within administrative area $A_i$ is 
\begin{align}
    C_{R} = \left[\sum_{\scaleto{\bmuu}{4pt} \in A_i} [2\pi R d(\bmuu, \bmt_{k})]^{-1}I({0<d(\bmuu, \bmt_{k}) < R})  \right]^{-1} \label{eq:ulocs:norm}
\end{align}

We take into consideration the different mechanism of geographical displacement for urban and rural clusters and we arrive at the following posterior for the true location 

\begin{equation}
    p(\bms = \bmt_k | \bmuu ,y ) \propto
    \begin{cases}
       N_k Q_k(2), & \text{if}\ y=1  
       \text{ (\emph{urban})}\\
      N_{k}  \left[ 0.99 Q_k(5)+ 0.01 Q_k(10) \right], &  \text{if}\ y=0  \text{ (\emph{rural})}
    \end{cases}
\end{equation}
where we define $Q_k(R^*)= \frac{C_{R=R^*}} {2R^* \pi d(u, t_k)}I({0 < d(\bmuu, \bmt_{k}) < R^*(km)})$
  
\begin{figure} [H]
    \centering
    \includegraphics[width=1\linewidth,trim={0 0cm 0cm 0},clip]{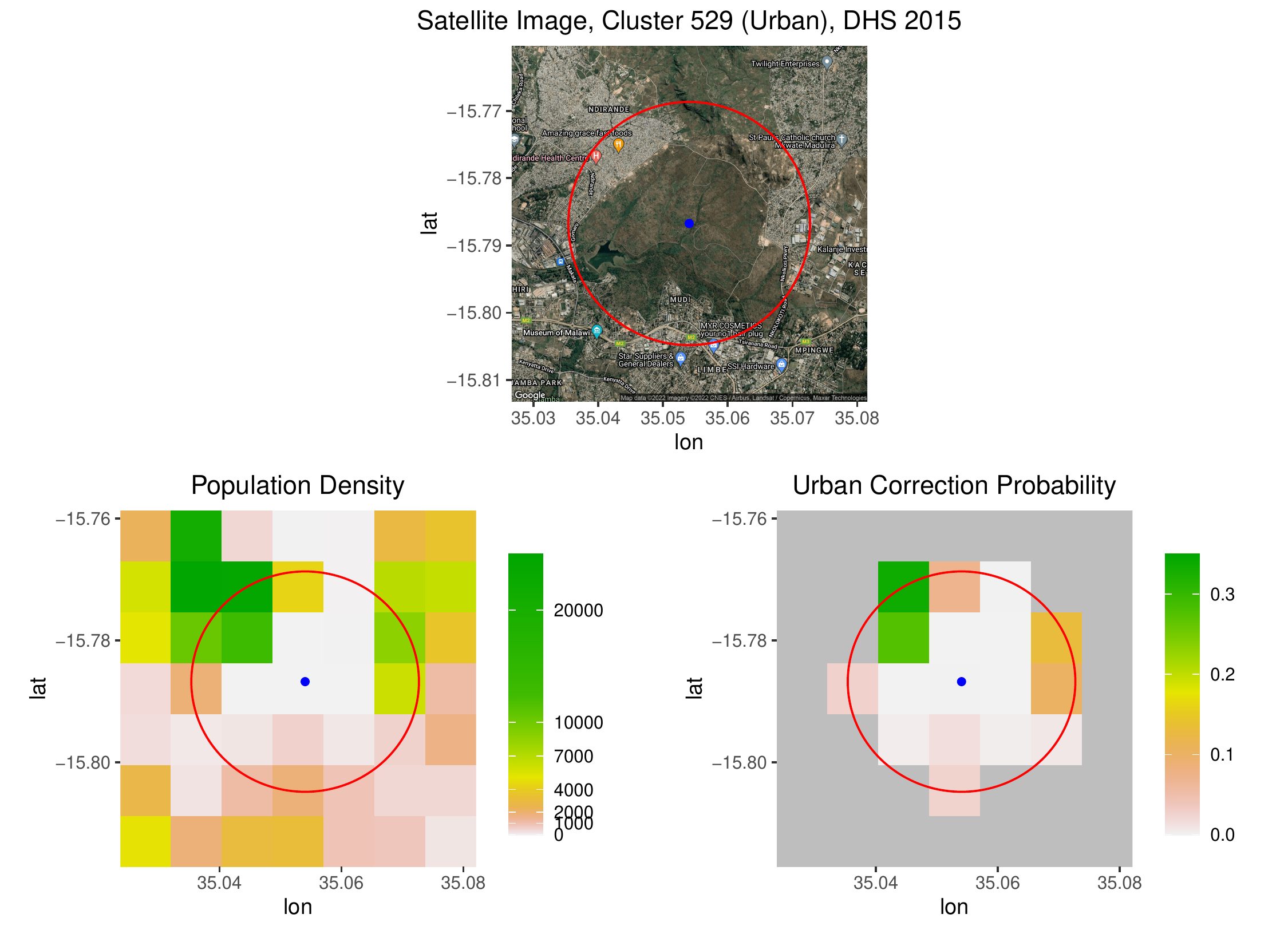}
     \caption{Example of a wrongly displaced urban cluster (cluster number 529 from Malawi DHS 2015-16). The blue dot is the observed cluster location and the red circle captures the potential true location for the cluster before jittering. The observed cluster location is in the middle of an unpopulated area, as shown from the terrain map and the population density map. }
    \label{fig:cluster_529}
\end{figure}

Roughly speaking, the posterior probability is proportional to population density and inversely proportional to the distance between candidate location and reported location. We note that this posterior is for a single cluster. By the independence of the geographical displacement for cluster locations, the joint posterior is the product of the posteriors for single locations, so that 
\begin{align}
    p(\{\bms_c\} | \{\bmuu_c\} ,\bmy ) = \prod_{c=1}^{c_0} p(\bms_c | \bmuu_c ,y_c )
\label{joint:loc}
\end{align}
where $c_0$ is the total number of clusters.

The lack of feedback from the classification model means that the partial adjustment will tend to relocate clusters towards high population density pixels. Such a tendency is not desirable for rural clusters, especially when the observed location for a rural cluster is close to an urban center. We adopt a practical approach in which we only adjust the locations for urban clusters, but leave the coordinates of rural clusters unchanged. In addition, instead of generating samples from $ p(\{\bms_c\} | \{\bmuu_c\} ,\bmy ) $ and fitting a classification model for each sample, we take the posterior mode from $ p(\{\bms_c\} | \{\bmuu_c\} ,\bmy )$ and fit the classification model once using the set of corrected values. The posterior mode is the most likely location that the cluster truly resides, under the model. As an example, the cluster shown in Figure \ref{fig:cluster_529} will be corrected to the pixel at the upper left corner, as its posterior probability is the highest among pixels in Figure \ref{fig:cluster_529}(c). See Figure \ref{fig:cluster_Blantyre} in the Appendix for a visualization of the implementation of partial correction in an entire admin-2 area.


\subsubsection{Full Bayesian Adjustment}
\label{full-adjust}
Now we model the true locations of clusters and the parameters from the classification model jointly. Such a scenario requires the classification model to also fall in the Bayesian framework. Candidate models include Bayesian logistic regression and BART.

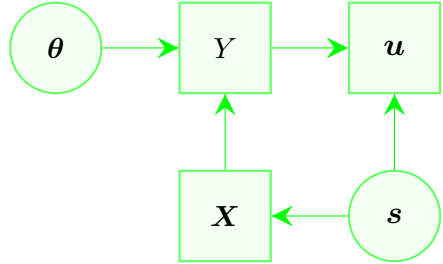
\begin{figure}[h]

\begin{center}

\begin{tikzpicture}[
roundnode/.style={circle, draw=green!60, fill=green!5,  thick, minimum size=12mm},
squarednode/.style={rectangle, draw=green!60, fill=green!5,  thick, minimum size=12mm},
]
\tikzset{bigarrow/.style={decoration={markings,mark=at position 1 with %
    {\arrow[scale=2.5,>=stealth]{>}}},postaction={decorate}}}
    
\node[squarednode]      (X)                              {$ \bmX$};
\node[squarednode]        (Y)       [above=of X] {$Y$};
\node[roundnode]      (theta)     [left=of Y]          {$ \bmtt$};
\node[roundnode]      (s)       [right=of X] {$\large \bms$};
\node[squarednode]        (u)       [right=of Y] {$ \large \bmuu$};

\draw[bigarrow,green] (s) -- (u);
\draw[bigarrow,green] (s) -- (X);
\draw[bigarrow,green] (X) -- (Y);
\draw[bigarrow,green] (Y) -- (u);
\draw[bigarrow,green] (theta) -- (Y);

\end{tikzpicture}

   \caption{Graphical model for the full Bayesian adjustment model.}

\label{fig:bayesian-graph}

     \end{center}
\end{figure}

The joint model can be decomposed as the following, based on the graphical model in Figure \ref{fig:bayesian-graph} 
\begin{equation}
p(\{\bms_c\},\bmtt,\bmy,\{\bmuu_c\})=p(\bmy|\{\bms_c\},\bmtt)\times p(\{\bmuu_c\}|\bmy,\{\bms_c\}) \times 
p(\{\bms_c\})\times p(\bmtt)
\end{equation}
The joint posterior then can be represented as

\begin{align}
\begin{split}
p(\{\bms_c\},\bmtt|\bmy,\{\bmuu_c\}) & = \frac{p(\{\bms_c\},\bmtt,\bmy,\{\bmuu_c\})}{p(\bmy,\{\bmuu_c\})}\\
       & \propto p(\bmy|\{\bms_c\},\bmtt)\times p(\{\bmuu_c\}|\bmy,\{\bms_c\}) \times 
p(\{\bms_c\})\times p(\bmtt)
\end{split}
\label{eq:joint-post}
\end{align}

To sample from the joint posterior, we propose a MCMC algorithm, which requires the conditionals $p(\bmtt|\bmy,\{\bms_c\},\{\bmuu_c\})$ and $p(\{\bms_c\}|\bmy,\bmtt,\{\bmuu_c\})$. Recall that $\bmtt$ are the parameters from the regression model, $\bmuu$ is the reported location and $\bms$ is the true location. 

The conditional posterior for $\bmtt$ is 
\begin{align}
\begin{split}
p(\bmtt|\bmy,\{\bms_c\},\{\bmuu_c\}) & \propto p(\bmy|\{\bms_c\},\bmtt)\times p(\bmtt)\\
       & = p(\bmy| \mathbf{X}_{\{\bms_c\}}, \bmtt) \times p(\bmtt)
\end{split}
\label{eq:cond-theta}
\end{align}

Sampling from this posterior is easily available from any classification model with a Bayesian setup, such as Bayesian logistic regression or BART. The model uses all training data such that the input consists of lists of labels, locations and covariates. Here we drop the conditioning on the reported locations because of the conditional independence shown in Figure \ref{fig:bayesian-graph}.

Based on Equation (\ref{eq:true-loc-post}), we write the conditional for $\{\bms_c\}$ as 
\begin{align}
\begin{split}
p(\{\bms_c\}|\bmy,\bmtt,\{\bmuu_c\}) & \propto p(\bmy|\{\bms_c\},\bmtt)\times p(\{\bmuu_c\}|\bmy,\{\bms_c\}) \times  p(\{\bms_c\})\\
       & = \prod_{c=1}^{c_0}| p(y_c|\bms_c,\bmtt)\times p(\bmuu_c|\bmy,\bms_c) \times  p(\bms_c) |\\
       & = \prod_{c=1}^{c_0}| p(y_c|\bmx_c,\bmtt)\times p(\bms_c|\bmuu_c,y_c)|
\end{split}
\label{eq:cond-s}
\end{align}

We observe that the joint conditional for all clusters is a product of the individual conditionals, which is expected as the jittering process is independent for different clusters. This also suggests that we can sample locations for individual clusters separately. Within the product on the last line of equation (\ref{eq:cond-s}), the first term can be viewed as a feedback from the classification model and the second term is the likelihood without considering the classification model. With both conditionals at hand, we shall use Gibbs sampling to sample from $p(\{\bms_c\},\bmtt|\bmy,\{\bmuu_c\})$. Details of the MCMC algorithm can be found in the Appendix.

\section{Results}
\label{sec:results}

\subsection{Classification Model Results}

We use Malawi as an example to demonstrate the performance of our various methods. The small island of Likoma which correspond to an admin-2 region is excluded, for reasons given in the Appendix. The goal is to produce U/R classifications for the sampling frame used in both DHS 2010 and DHS 2015--2016. 

To reiterate, the cluster U/R status and the associated GPS coordinates serve as training data (with 1655 clusters in total). In addition, we include the geographical covariates listed in Table  \ref{table:cov_summary} for the classification models. The methods we use, unless otherwise noted, are all benchmarked to the known admin-2 urban fractions from the census.

The models we consider are listed as follows,
\begin{enumerate}
\item
Population Thresholding: no statistical model, directly threshold using total population density to meet benchmark.
\item
BART: three versions considered. Using original cluster locations (BART-uncorrected), corrected cluster locations (BART-corrected, method from Section \ref{Partial-Adjustment}), and MCMC algorithm (BART-MCMC, from Section \ref{full-adjust}). 
\item 
Bayesian logistic regression (Logistic Regression).
\item 
Gradient Boosted Trees (GBT).
\end{enumerate}

Across all models, population density and night time light are the most important predictors. In logistic regression model, they are the only two covariates that are significant (under $\alpha=0.05$).

\begin{figure}[H]
    \centering
    \makebox{
    \includegraphics[width=0.9\linewidth,trim={0 0cm 0 1cm},clip]{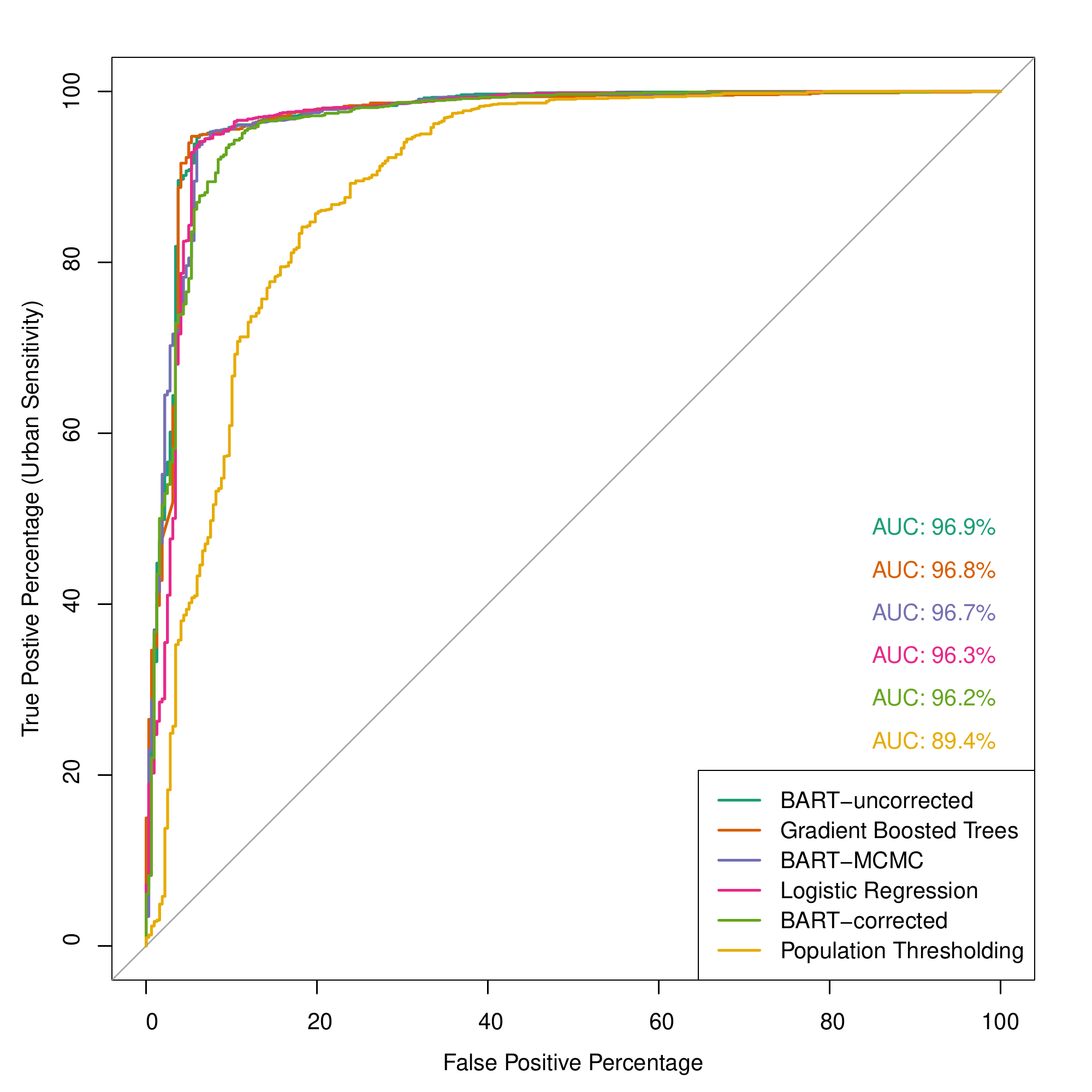}
    }
    \caption{ROC curve for 5-fold CV across methods.}
        \label{fig:ROC-curves}
\end{figure}

\subsubsection{Cross Validation Results}

We first examine the classification accuracy of pixels using cross validation. The clusters are split into 5 folds and classifications from the testing sets are aggregated to form a single ROC curve. In the testing set, the original cluster locations and associated covariate values are used for prediction. Hence, the observed U/R status in the testing set might not reflect the truth due to the geographical displacement process described in Section \ref{sec:jitter}. However, the ROC curves do serve as a fair comparison for the performance of the methods. The results are shown in Figure \ref{fig:ROC-curves}, where we observe that the naive population thresholding method is outperformed by all other methods. The statistical classification models share comparable performance as further shown by the zoomed ROC curves in Figure \ref{fig:ROC-curves-zoomed} in the Appendix. 

\subsubsection{Pixel Classification Visualization}

The motivation for our method is the absence of an U/R classification map in countries in which we wish to carry out SAE of health and demographic indicators in. Hence, we do not have the ground truth for checking the pixel-level classification we produce. However, we might use a surrogate such as the google (stamen) maps to visually validate our classification results. Admittedly, such comparison is subjective and lacking of statistical rigour, but it provides further evidence of the performance of the methods we propose. 

\begin{figure}[H]
    \centering
    \includegraphics[width=0.95\linewidth,trim={0 0cm 0 0cm},clip]{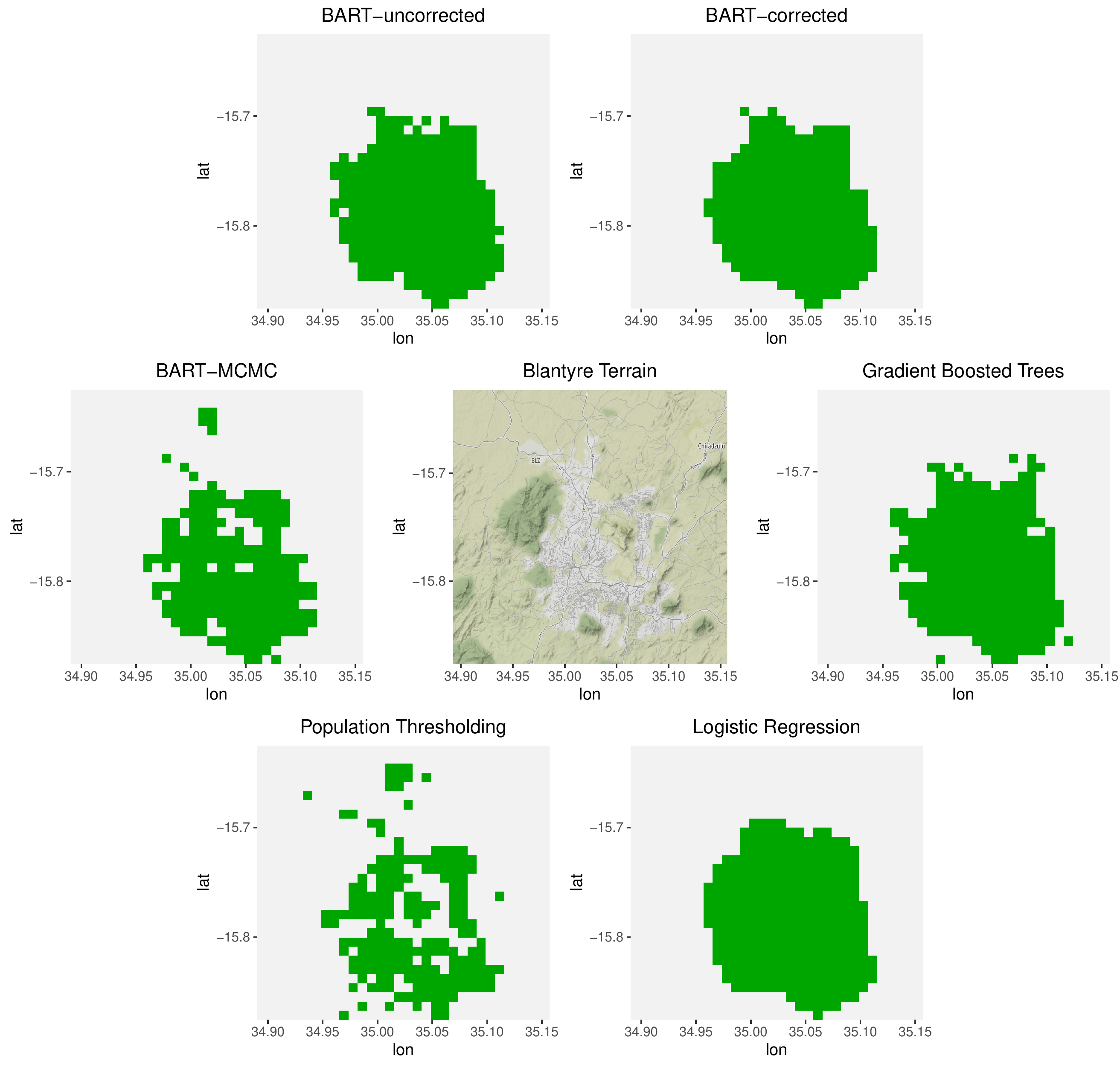}
    \caption{Classifications from different models for admin-2 region Blantyre in Malawi. Green pixels represent the areas predicted as urban.}
        \label{fig:Blantyre-Classification}
\end{figure}

We take as an example the urban area in Blantyre, which is the biggest city in Malawi. The plot in the center of Figure \ref{fig:Blantyre-Classification} shows the terrain map for part of the admin-2 region of Blantyre. A complete map for this region can be found as Figure \ref{fig:Blantyre-map} in the Appendix. The urban area is distinguished by the white color and the U/R delimitation is rather clear because Blantyre city is surrounded by mountains.
Note that the map here is contemporary while the U/R classification we aim to construct is for the 2008 census sampling frame. The fairness of comparison relies on a slow expansion rate for Blantyre city \citep{gondwe2021analysis}.

Figure \ref{fig:Blantyre-Classification} compares the U/R classification in Blantyre using the models we proposed. All methods recover the approximate shape of the urban regions, but they possess different degree of fuzziness at the boundaries. The oversmoothed shape resulting from logistic regression might be attributed to the smoothness in the covariates surface and the linearity that logistic regression imposes on the logit scale. The tree-based algorithms performs better at depicting the boundaries and BART-MCMC is arguably the best method as it even recovers the unpopulated mountain area within the center of the urban area. The non-statistical method, population thresholding, yields reasonable results, though the realized urban area is too diffuse.

The performance for population thresholding might not be generalized to other admin-2 areas because Blantyre is highly urban (the most urban region in Malawi) and population density has high predictive power here. On the other hand, the rest of the methods incorporate additional covariates and are more robust, so we might expect them to yield stable performance across admin-2 regions.

\begin{figure}
    \centering
    \includegraphics[width=0.95\linewidth,trim={0 0cm 0 0cm},clip]{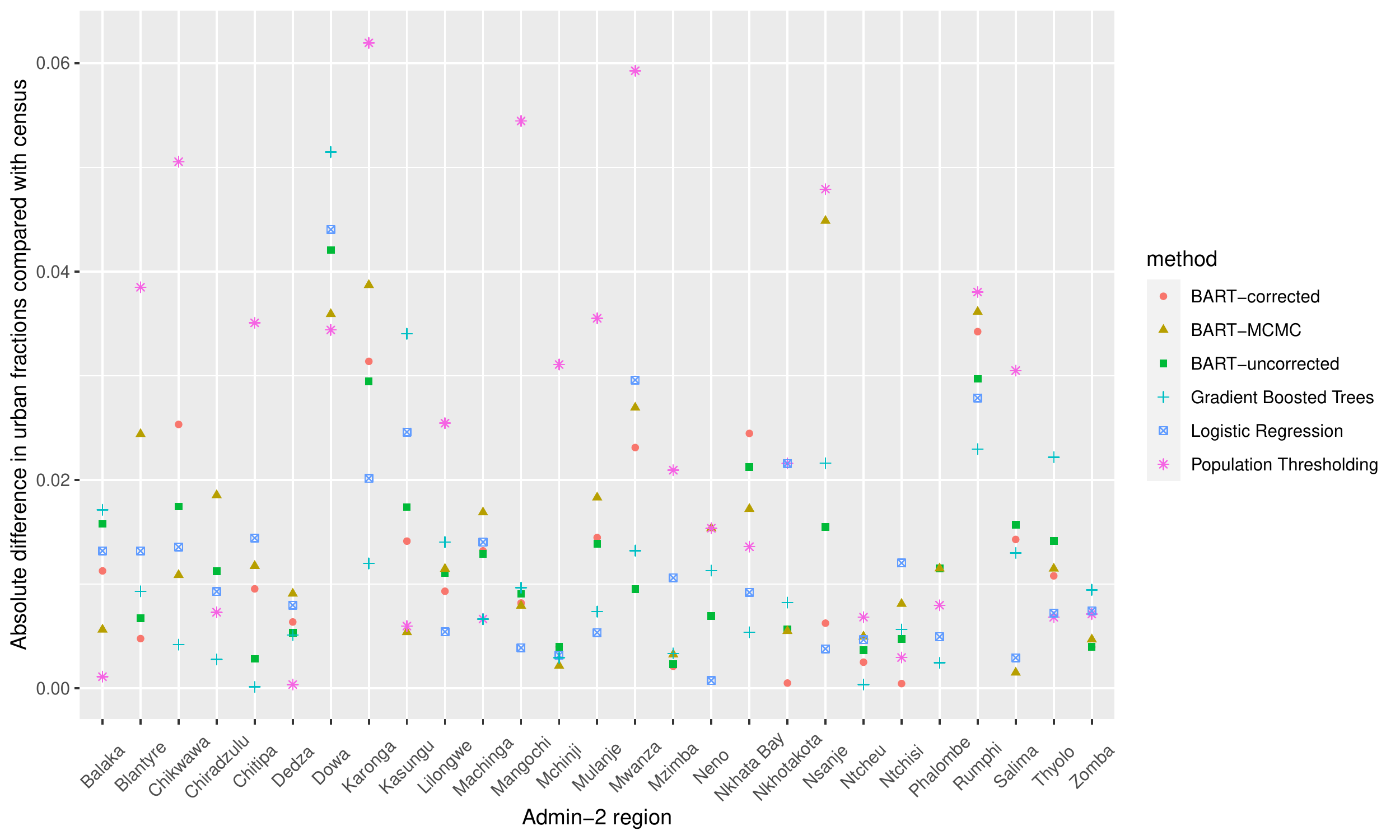}
        \caption{Benchmark at national level and comparison of the resulting admin-2 urban fractions with truth.}
    \label{fig:malawi-frac-check}
\end{figure}

\subsubsection{Urban Fraction Accuracy}
\label{sec:urb-frac-accuracy}
\begin{table}
\caption{\label{table:admin2-error}National thresholding, admin-2 proportions error}
\centering
\begin{tabular}{rlrr}
  \hline
 & Method & Mean Error& Mean Absolute Error \\ 
  \hline
& BART-corrected & -0.26\% & 1.28\% \\ 
   & BART-MCMC & -0.34\% & 1.46\% \\ 
  & BART-uncorrected & -0.12\% & 1.27\% \\ 
  & Gradient Boosted Trees & 0.02\% & 1.17\% \\ 
  & Logistic Regression & -0.24\% & 1.24\% \\ 
  & Population Thresholding & -0.03\% & 2.47\% \\ 
   \hline
\end{tabular}
\end{table}
Checking classification map for pixels is visually appealing, but admittedly, such comparison is more subjective and less quantifiable. Considering the ultimate goal is to accurately estimate the urban population fraction in a certain area, we present results comparing the performance of methods in estimating urban population fractions. 

From the 2008 census in Malawi, we can obtain the U/R specific population within each admin-2 region. In the complete estimation procedure, we can calculate the corresponding admin-2 urban population fractions and use them as benchmarks. To check the accuracy of our methods, we will instead use national level urban fraction for benchmarking. Then we will use the resulting classification map to calculate urban fractions at admin-2 level and compare with the truth from the census. Such treatment also mimics the situation in most other countries, as admin-2 level urban fractions are not generally available. Note that although we choose total population for quality checking purpose only, our method can be applied to any sub-population depending on the availability of population density surfaces. 

We measure the errors at the absolute scale, rather than the relative scale, because the aggregation for prevalence is a linear combination of urban and rural estimates. Figure \ref{fig:malawi-frac-check} shows the absolute difference between estimated urban fractions with the true fractions. Table \ref{table:admin2-error} summarizes the mean errors and mean absolute errors for the comparison, where the errors are calculated as the difference between estimated urban fractions and the known urban fractions from the census, for the total population; the errors are averaged across 27 admin-2 regions in Malawi (Likoma excluded). The methods using statistical models for classification all perform better than the naive population thresholding method. Gradient boosted trees outperforms logistic regression and versions of BART by a small margin. The overall results are satisfactory as the difference are measured at absolute scale, i.e., on average we only over or under estimate the true urban fraction by $\sim$1.2\%.

\subsection{Data Application: HIV Prevalence -- Space-only Modeling}
\label{sec:HIV-example}
We apply our method to estimate the HIV prevalence for women aged 15-49 in Malawi in 2015. In addition to the BYM2 spatial model described in Section \ref{prev-model}, we consider two additional ways to model the area-level effects, namely a fixed effects model and an iid random effects model. The details of the model setup can be found in the Appendix. We stress that Malawi is not typical in the availability of data, in particular, in richness -- for most countries, there would be insufficient data to fit the fixed effects models.

U/R specific estimates are aggregated using urban proportion for female 15-49 population for Malawi in 2015. Classification surface is obtained from BART with uncorrected cluster locations, benchmarked to known admin-2 urban proportion for total population. For Likoma, we use the urban fraction for the total population as an replacement, for reasons described in the Appendix.

First, to illustrate the association between HIV prevalence and U/R, we present the odds ratio corresponds to urban effect in the fixed effect model with U/R intercept. The estimate for odds ratio (95\% CI) is 2.03 (1.69, 2.44), indicating that the odds of having HIV is 2.03 times higher among women in urban areas as compared to rural areas.

We compare estimates from various models with survey weighted direct estimates for HIV prevalence using all available data, i.e., not cross validated. Table \ref{table:HIV-est-res} presents results for three classes of cluster-level models: fixed effects models, area-level IID random effects models and spatial BYM2 models. In addition, we consider three stratification methods within each class of model, namely no U/R stratification (Area only), U/R intercept (U/R+ Area) and full U/R adjustment (U/R $\times$ Area).

We consider a number of metrics for evaluating model performance by quantifying the difference between model estimates and survey weighted direct estimates. Given point estimates from the model, $\widehat{\bp}=(\widehat{p}_1, \dots ,\widehat{p}_m)^{\text{\tiny{T}}}$,
and direct estimates $\tilde{\bp}=(\tilde{p}_1, \dots ,\tilde{p}_m)$ for the $n_{\scaleto{A}{4pt}}=28$ admin-2 areas, we calculate the overall scoring metrics for a model as,
\begin{eqnarray}
\mbox{Bias}(\widehat{\bp}, \tilde{\bp}) &=& \frac{1}{n_{\scaleto{A}{4pt}}} \sum_{i=1}^{n_{\scaleto{A}{3pt}}} 100 \times (\widehat{p}_i - \tilde{p}_i) \label{eq:bias} \\
\mbox{Absolute Bias}(\widehat{\bp}, \tilde{\bp}) &=& \frac{1}{n_{\scaleto{A}{4pt}}} \sum_{i=1}^{n_{\scaleto{A}{3pt}}} 100 \times |\widehat{p}_i - \tilde{p}_i| \label{eq:absolutebias} \\
\mbox{Relative Bias}(\widehat{\bp}, \tilde{\bp}) &= &\frac{1}{n_{\scaleto{A}{4pt}}} \sum_{i=1}^{n_{\scaleto{A}{3pt}}} 100\% \times \frac{\widehat{p}_i - \tilde{p}_i}{\tilde{p}_i} \label{eq:cv1}\\
\mbox{Absolute Relative Bias}(\widehat{\bp}, \tilde{\bp}) &= &\frac{1}{n_{\scaleto{A}{4pt}}} \sum_{i=1}^{n_{\scaleto{A}{3pt}}} 100\% \times \left|\frac{\widehat{p}_i - \tilde{p}_i}{\tilde{p}_i}\right| \label{eq:cv1}
\end{eqnarray}

\begin{table}
\caption{\label{table:HIV-est-res} Model comparison for HIV prevalence of women of aged 15-49 in Malawi.}
\centering
\begin{tabular}{llrrrrr}
  \hline
  Model type & Stratification  & Bias & \makecell[l]{Absolute \\ bias} & \makecell[l]{Relative \\ bias}&
  \makecell[l]{Absolute \\relative bias} & \makecell[l]{CI \\width} \\ 
  \hline
 Fixed effects & Area only  & 0.75 & 0.86 & 9.47\% & 10.01\% & 7.75 \\ 
  Fixed effects & U/R+Area  & -0.03 & 0.57 & 0.27\% & 5.45\% & 7.09 \\ 
Fixed effects & U/R*Area & 0.14 & 0.41 & 0.74\% & 4.97\% & 7.53 \\ 
  \hline
  Area IID  & Area only & 0.84 & 1.25 & 15.23\% & 17.22\% & 7.48 \\ 
  Area IID  & U/R+Area & 0.22 & 0.89 & 7.27\% & 11.01\% & 6.90 \\ 
  Area IID  & U/R*Area & -0.17 & 0.70 & 2.79\% & 8.42\% & 7.01 \\ 
    \hline
  BYM2 & Area only & 0.37 & 1.19 & 9.90\% & 14.60\% & 6.53 \\ 
 BYM2 & U/R+Area & -0.23 & 0.99 & 2.35\% & 11.02\% & 5.94 \\ 
  BYM2 & U/R*Area  & -0.25 & 0.96 & 1.27\% & 11.61\% & 6.16 \\ 
   \hline
\end{tabular}
\end{table}

Table \ref{table:HIV-est-res} summarizes the model performance via the above scoring rules. The values for bias and absolute bias are on the per 100 people scale. For example, on average the fixed effects model without U/R stratification overestimate HIV prevalence by 0.75 per 100 women, compared with the direct estimates. We expect such bias because urban areas have higher HIV prevalence and oversampling of urban results in overestimates of HIV prevalence.

\begin{figure} [H]
    \centering
    \makebox{
    \includegraphics[width=0.95\linewidth,trim={0 3cm 0 2cm},clip]{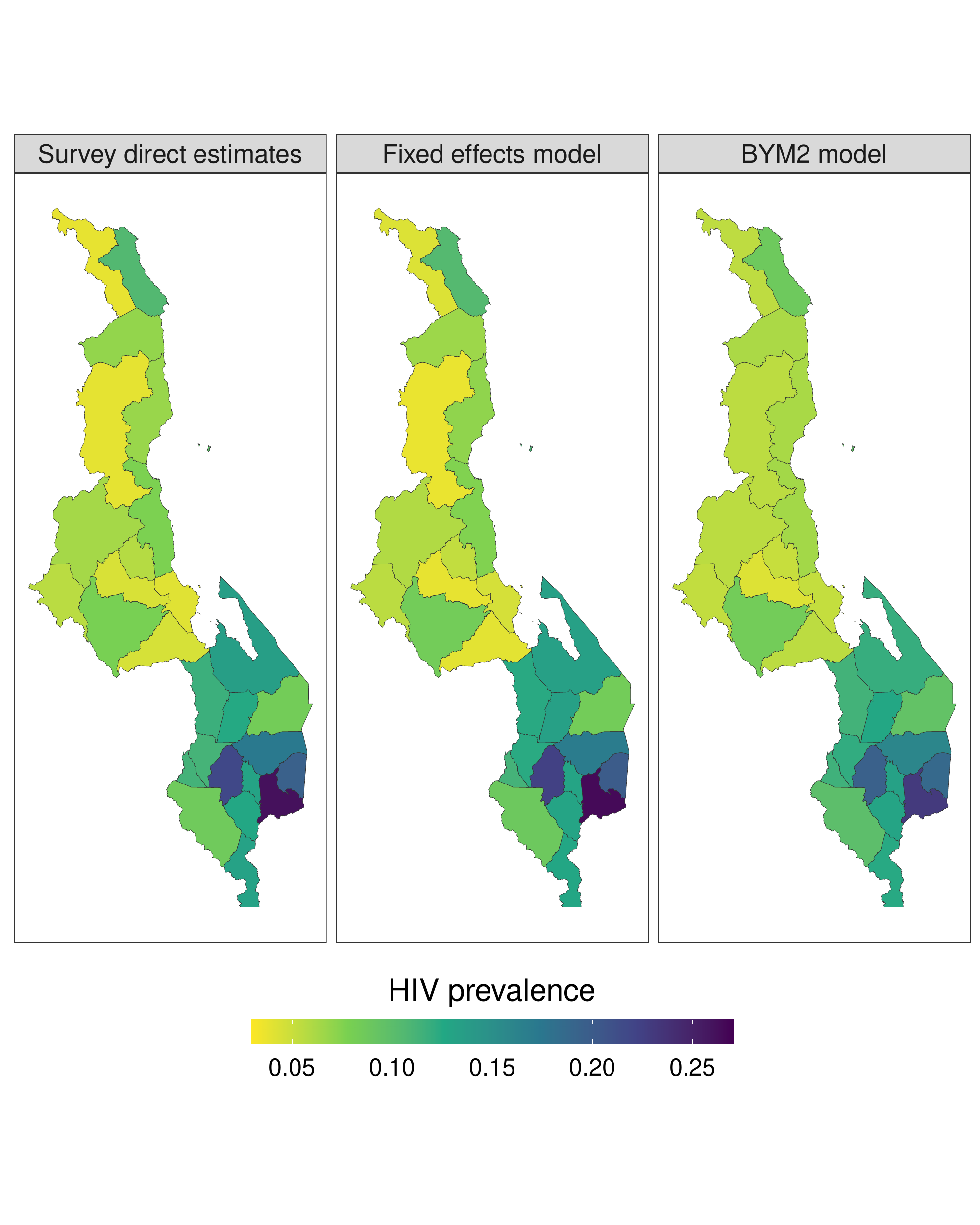}
    }
    \caption{Maps for HIV prevalence estimates at admin-2 level from three models.}
        \label{fig:HIV-admin2-map}
\end{figure}

For all types of models, U/R stratification significantly reduces the bias. The fixed effects models are not affected by area level smoothing and thus are not susceptible to shrinkage, which introduces bias. Focusing on the first three row in Table \ref{table:HIV-est-res}, we observe that the absence of U/R stratification leads to overestimation of HIV prevalence by 9.47\% and U/R adjustments limit the relative bias to be below 1\%. Let us revisit the motivating plots presented in the introduction section, where we compare estimates from fixed effect models with and without U/R stratification (row 1 and row 3 in Table \ref{table:HIV-est-res}). From the comparison between the two panels in Figure \ref{fig:stratified-compare-2}, we can see that the bias disappears when we include U/R stratification in the cluster-level model. Figure \ref{fig:HIV-admin2-map} maps the admin-2 estimates from survey direct estimator, fixed effects model and BYM2 model with full U/R adjustment (row 3 and row 9). Spatial smoothing can be observed in the BYM2 model.

For point estimates, across all type of models, full U/R adjustment (U/R*Area) is slightly favored over U/R intercept (U/R+ Area) as they produce estimates closer to the direct estimates. As a trade-off for flexibility, the full U/R adjustment estimates have higher uncertainty compared with the U/R intercept counterparts, since more parameters are being estimated. To sum up, models with U/R stratification  account for the survey design and ignoring U/R stratification leads to bias in model estimates. In general, it may not be necessary to consider the full U/R*Area model. Spatial smoothing model (BYM2) could also yield reliable estimates at finer resolution. In the Appendix, we show the results for admin-3 level HIV prevalence in Malawi from BYM2 models with and without U/R stratification.

\section{Discussion}
\label{sec:discussion}

In this paper we have proposed a framework to account for stratification in unit-level models. In particular, we deal with the U/R stratification in DHS and other household surveys. Obtaining strata specific estimates from unit-level models is relatively straightforward, as the U/R status for clusters are available and can easily be included as model terms. The major contribution of the paper is to establish a comprehensive aggregation procedure for obtaining the U/R population fraction for aggregation. 

The classification methods we proposed in Section \ref{sec:modeling} significantly outperform the naive population thresholding method used in \cite{wu:etal:21}. Among the methods based on statistical models, BART and GBT with original cluster location yield estimates closest to the truth, as shown in Section \ref{sec:urb-frac-accuracy}. We might have expected the BART-MCMC and BART-corrected method to work better as they properly account for the jittering process from DHS surveys, but their performance was slightly interior to the methods that used the original cluster locations. Such results might well be country specific as the urban fractions across admin-2 regions in Malawi vary greatly. In terms of computation time, MCMC-BART is much more demanding than the other methods. Therefore, based on our current experience, for now we recommend using BART and GBT methods with the original cluster location. Logistic regression can also serve as an alternative, for its fast computation and accuracy in estimating urban fractions, but it is less flexible in handling interactions between covariates. We plan to apply our method in other countries and with other indicators and a comparison of methods across more countries will provide more concrete guidance. 

Removing bias caused by the association between a health indicator of interest and U/R is the overarching goal for incorporating U/R stratification in a cluster-level model. Through the example of modeling HIV prevalence for women of age 15-49 in Malawi, we demonstrate that the proposed methods achieved such a goal. If we treat the estimated urban fractions as fixed values, the aggregated estimates tend to have smaller uncertainty compared with the estimates without U/R stratification. This observation is consistent with the general idea that stratified sampling leads to survey estimates with reduced variance.

\bibliographystyle{rss} 
\bibliography{refs}

\newpage
\appendix

\section{Appendix}
\label{sec:appendix}

\subsection{MCMC Algorithm for BART-MCMC}

The MCMC algorithm is:
\begin{enumerate}
   \item Initialize $\bmtt^{(0)}=B(\{\bms_c\},\bmy, \{\bmx_{\bms^{(0)}_c}\})$, where the location for cluster $c$, $\bms^{(0)}_c$ is sampled via $p(\bms_c|\bmuu_c,y_c)$ and $B$ represents the classification model, such as BART or Bayesian logistic regression.
   \item Iterate:
   \begin{enumerate}
     \item Sample updated location $\bms_c^{(m+1)}$ for each single cluster $c$ using Gibbs sampling,
\begin{equation}
 p(\bms_c|y_c, \bmtt^{(m)},\bmuu_c)  \propto p(y_c|\bms_c,  \bmtt^{(m)})  \times  p(\bms_c |\bmuu_c,y_c) 
\end{equation}
Note that each cluster could be updated individually as the jittering process is independent. As derived in equation (\ref{eq:true-loc-post}), the first term on the right is the Bernoulli predicted probability that pixel at location $\bms_c$ has the same U/R status as observed, i.e.,
\begin{equation}
p(y_c|\bms_c,   \bmtt^{(m)}) =v_c(\bmtt^{(m)})^{I(y_c=urban)}\cdot (1-v_c(\bmtt^{(m)}))^{I(y_c=rural)}
\end{equation}
where $v_c(\bmtt^{(m)})$ is the predicted probability of pixel at location $\bms_c$ is urban using a single set of posterior draws $\bmtt^{(m)}$.\\

     \item Calculate predicted probabilities $v_g(\bmtt^{(m)})$ on national grid with pixels indexed by $g$.\\

     \item Fitting BART to get prosterior of tree parameters. Update \\ $\bmtt^{(m+1)}=B(\{\bms_c^{(m+1)}\},\bmy,\{\bmx_{\bms^{(m+1)}_c}\})$
     
   \end{enumerate}

\item Summarize: For point estimates, we follow equation (\ref{eq:tilde_mu_g}), (\ref{eq:tilde_tau}) and  (\ref{eq:tilde_c_g}) to calculate the U/R predicted probability surface $\tilde{\bmmu}=\{\tilde{\mu}_g\}$, area specific $\{\tilde{\tau}_i\}$ and classifications $\{\tilde{C}_g\}$. Since from the MCMC algorithm, draws of $\bmtt^{(m)}$ and $\{v_g(\bmtt^{(m)})\}$ are available, it is possible to report the uncertainty of estimated urban fractions. We outline the procedure for quantifying uncertainty in the Appendix.  

 \end{enumerate}

\subsection{Under-5 Mortality (U5MR) -- Space-time Modeling}

\begin{figure}[H]
    \centering
    \includegraphics[width=0.95\linewidth,trim={0 0cm 0 0cm},clip]{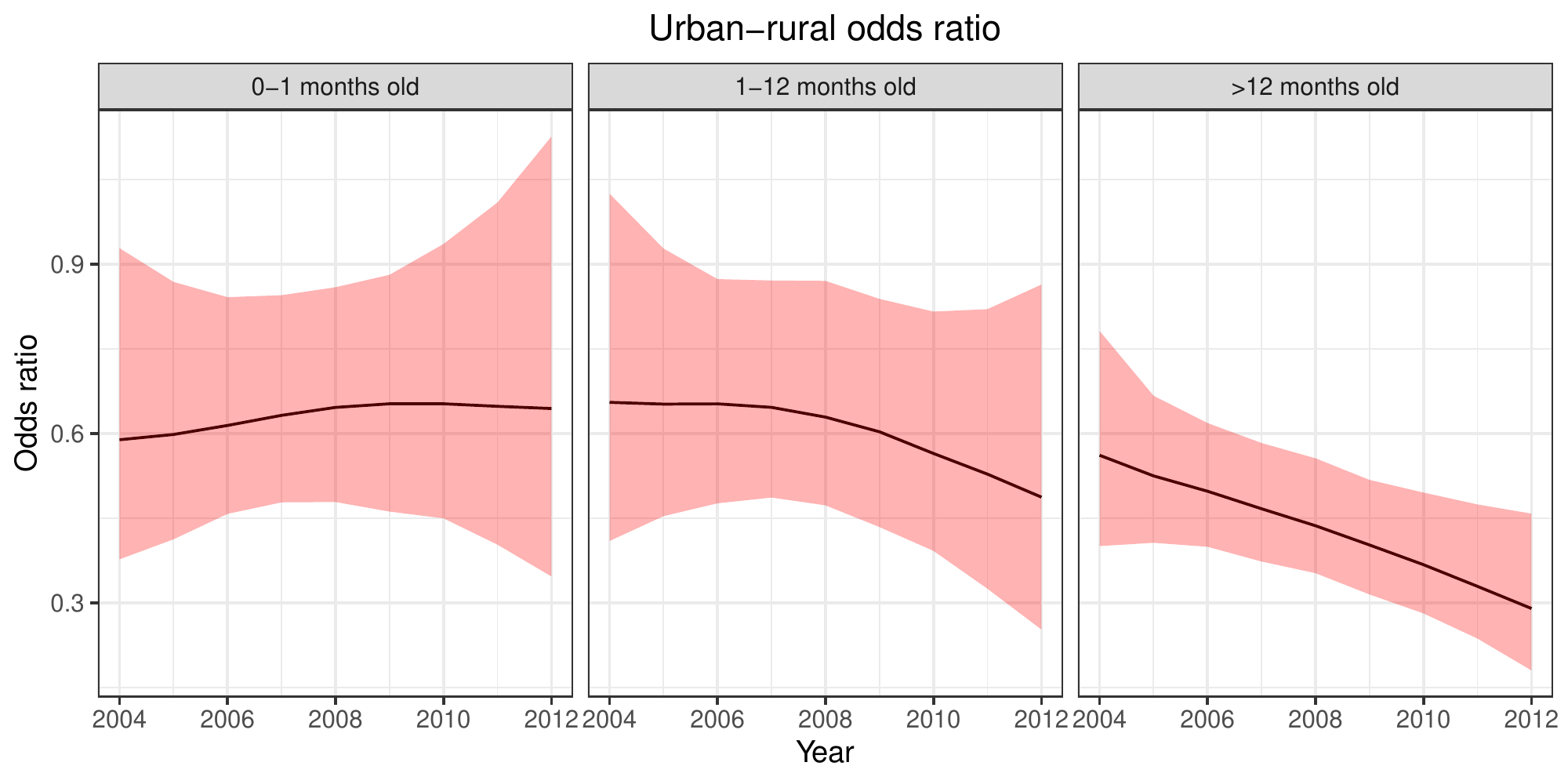}
        \caption{Odds ratios for mortality (urban/rual) over time for 3 groups of age bands. }
    \label{fig:Niger-3UR-odds-ratio}
\end{figure}

We apply our method to estimate U5MR in Niger. We use Niger DHS survey conducted in 2012 to produce subnational U5MR estimates from 2004-2012. U5MR in Niger displays a big U/R discrepancy. Figure \ref{fig:Niger-3UR-odds-ratio} shows the odds ratio of mortality comparing urban over rural, indicating that rural mortality is much higher than urban mortality for all age groups.
In addition, Niger DHS 2012 oversampled urban areas in most of the admin-1 regions. Thus, it is important to account for the U/R stratification in modeling.

\subsubsection{Prevalence Model}
\label{sec:u5mr-prev}
We adopt the same modeling strategy as in \cite{wu:etal:21}. Here we only present a brief overview of the modeling choices, we refer the readers to \cite{wu:etal:21} for further details. We model the U5MR as discrete hazards and assume a beta-binomial model for the probability (hazard)
of death from month $m$ to $m+1$ at cluster level. We assume constant hazards
within each age bands such that the number of deaths occurring within age
band $a[m]$, in cluster $c$ and time $t$, follow the
beta-binomial distribution, 
\begin{equation}\label{eq:BB1-age}
Y_{a[m],c,t} ~|~  h_{a[m],c,t} ,d\sim \mbox{BetaBinomial}\left(~ n_{a[m],c,t},~ h_{a[m],c,t},~d~
\right),
\end{equation} where $h_{a[m],c,t}$ is the monthly hazard for age
band $a[m]$, in cluster $c$, time $t$, and $d$
is the overdispersion parameter. 

The general modeling strategy is to treat the hazards as a function of space, time, child's age and U/R strata. To acknowledge the sampling design, we include separate urban and rural temporal terms. Sparsity of data lead us to several parsimonious choices. We assume a same U/R association across all areas. We also hold U/R associations constant within in each of the three groups of age bands. We allow U/R association to vary over time, but smooth it via RW2s.

The hierarchical model for hazard can be further specified as,
\begin{eqnarray}
h_{m,c,t} &=& \mbox{expit}( \eta_{m,c,t} +   \tau_t ) \label{eq:BB2-age}\\
\eta_{m,c,t} &=& \alpha_{a[m]}+
\alpha^\star_{a^\star[m],t} \times I(\bolds_c \in \mbox{ rural }) \nonumber\\&+& (\gamma_{a^\star[m]} +
\gamma^\star_{a^\star [m],t} )\times I(\bolds_c \in \mbox{ urban })\nonumber\\
& +&  
b_{i[\boldsymbol{s}_c]}  +\delta_{i[\boldsymbol{s}_c],t} 
\label{eq:BB3-age}
\end{eqnarray}
The above formulation sets $\alpha_{a[m]}$ and $\gamma_{a^\star[m]} $ U/R specific age-band main effects, with $\alpha^\star_{a^\star[m],t}$ and $\gamma^\star_{a^\star [m],t} $ as their temporal smoothers. We include time-invariant spatial effects as $b_{i[\boldsymbol{s}_c]}$, and space-time interactions as $\delta_{i,t}$. 

For a given area $i$ and time $t$, the estimated strata U/R specific U5MRs are,

\begin{eqnarray}\label{eq:U5MR}
\mbox{U5MR}_{i,t,R} &=& 
1- \prod_{a=1}^6
\left[ \frac{1}{1+\exp(
\alpha_{a}  + \alpha^\star_{a,t}  + b_{i}  +\delta_{i,t}) }\right]^{z[a]}\\
 \mbox{U5MR}_{i,t,U} &=& 
1- \prod_{a=1}^6
\left[ \frac{1}{1+\exp(\alpha_{a} +
\gamma^\star_{a} + \gamma^\star_{a,t} +b_{i}  +\delta_{i,t}) }\right]^{z[a]},
\end{eqnarray} 
where \(z[a]=1,11,12,12,12,12\), are the default number of months within each of the six age bands. 

\subsubsection{Aggregation Model}

Similar to the space-only setting, the temporal extension for the aggregation model can be formulated as,
\begin{eqnarray}
\mbox{U5MR}_{i,t} &=& \mbox{U5MR}_{i,t,U}\times q_{i,t} + \mbox{U5MR}_{i,t,R}\times (1-q_{i,t})
\end{eqnarray}
where $q_{i,t}$ and $1-q_{i,t}$ are the proportions
of the under-5 population in area $i$ that are urban and rural in year
$t$. Note that the urban fractions also vary with time $t$.

\begin{figure} [H]
    \centering
    \makebox{
    \includegraphics[width=0.95\linewidth,trim={0 3cm 0 3cm},clip]{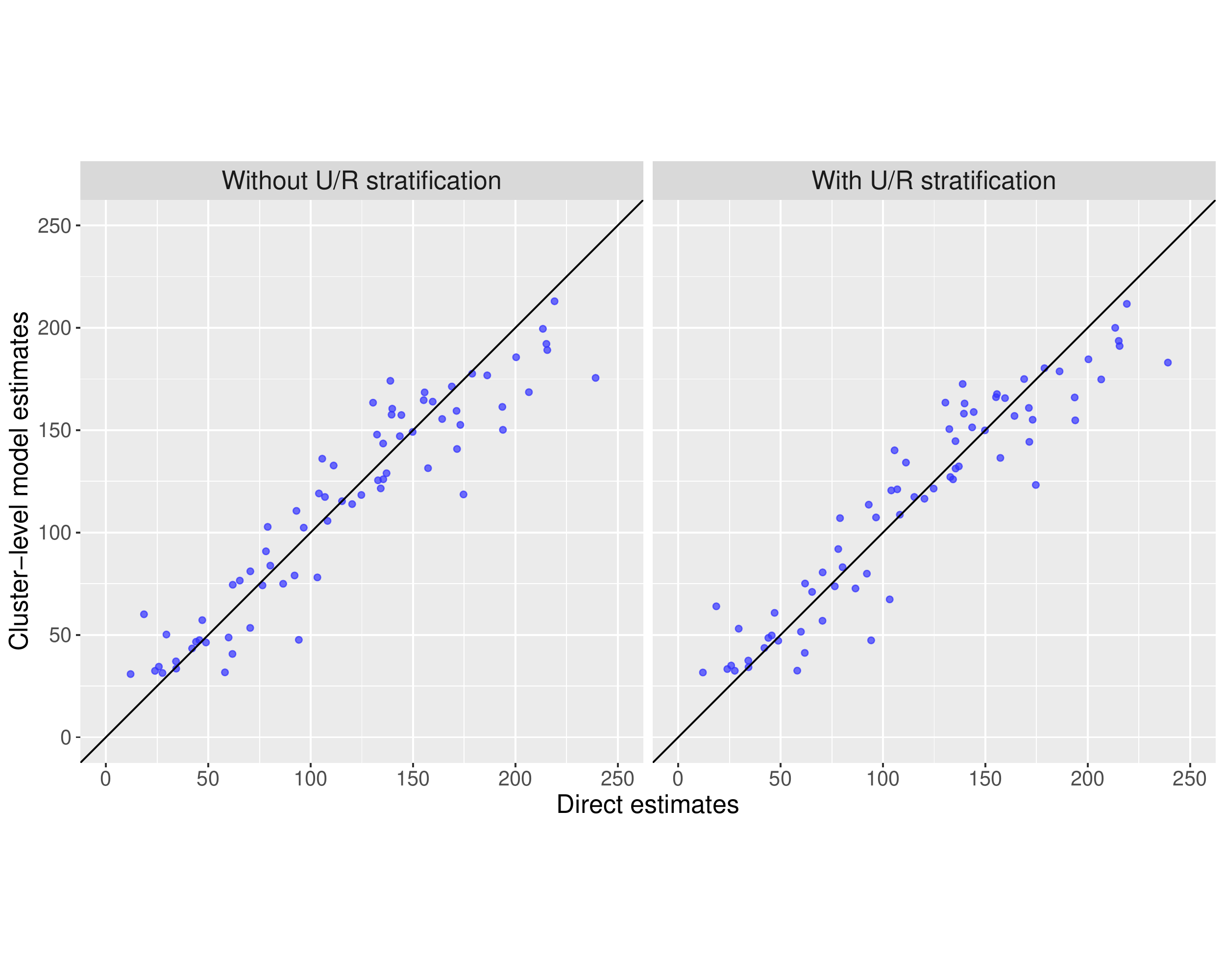}
    }
    \caption{Comparison of cluster-level model estimates with direct estimates for U5MR in admin-1 regions of Niger over 2004-2012. Numbers displayed are deaths per 1000 live births. }
        \label{fig:u5mr-admin1-all-year}
\end{figure} 
\subsubsection{Classification and Fraction Model}

We adopt BART with original cluster location as the classification model, benchmarked to known admin-1 urban proportions for total population. After obtaining the estimated U/R classification surface $\{\tilde{C_g}\}$, we can extend equation (\ref{eq:classify-to-frac-est}) to calculate year specific urban fractions for area $A_i$ in year $t$,
\begin{equation}
\tilde{q}_{i,t} = \frac{\sum_{g \in A_i} I(\tilde{C}_g = \text{urban}) \times L_{t,g}}{\sum_{g \in A_i} L_{t,g}}
\end{equation}
where $g$ is the index for the pixel and $L_{t}$ is the under-5 population surface in Niger at year $t$. The estimated fractions $\tilde{q}_{i,t}$ can then be used in the aggregation model.

We emphasize that U/R classification surface remain constant and the variation of urban fractions throughout time solely comes from the change in population density.

\begin{figure} [H]
    \centering
    \makebox{
    \includegraphics[width=0.95\linewidth,trim={0 0cm 0 1cm},clip]{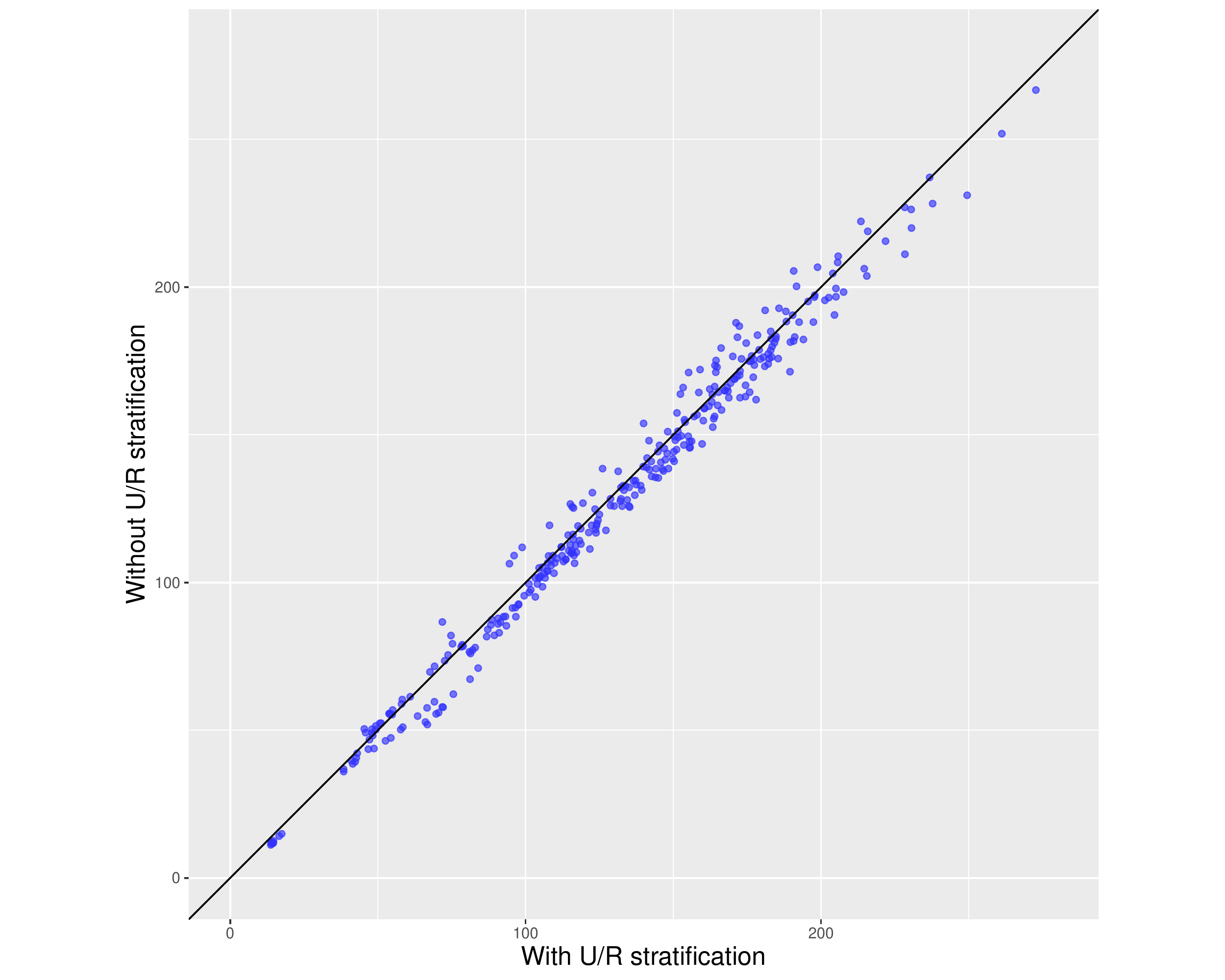}
    }
    \caption{Comparison of cluster-level model estimates with direct estimates for U5MR in admin-1 regions of Niger over 2004-2012. Numbers displayed are deaths per 1000 live births. }
        \label{fig:u5mr-admin2-all-year}
\end{figure}
\subsubsection{Results}

Figure \ref{fig:u5mr-admin1-all-year} compares cluster-level model estimates with survey weighted direct estimates using models described in section \ref{sec:u5mr-prev}. Each dot represents estimate for one admin-1 region at a single year between 2004-2012 and left/right panels are from model without/with U/R stratification. We observe that estimates for regions with high U5MR (upper right of each panels) are pulled closer to the direct estimates. This evidence suggests that adjusting for U/R stratification corrects the under-estimation of U5MR caused by oversampling of urban clusters in less developed regions.

\begin{figure} [H]
    \centering
    \makebox{
    \includegraphics[width=0.95\linewidth,trim={0 2cm 0 0cm},clip]{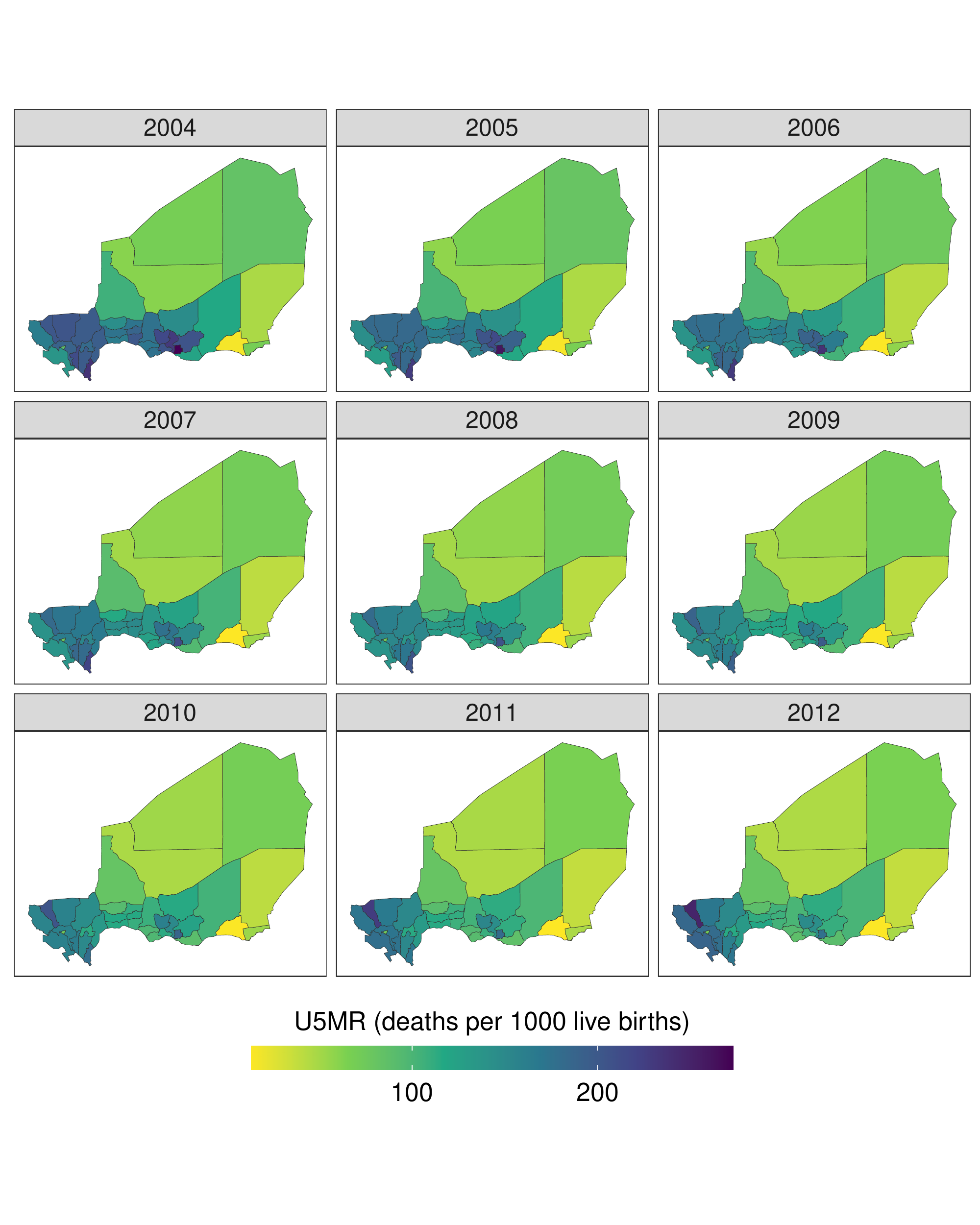}
    }
    \caption{U5MR estimates from cluster-level model with U/R stratification estimates, Niger admin-2, 2004-2012}
        \label{fig:u5mr-admin2-map}
\end{figure}

On average, the model without U/R stratification underestimate U5MR by 2.6 per 1000 births while the U/R stratified model reduce the bias to 0.9 per 1000 births. The reduction is more significant in the 6 less developed admin-1 areas (urban population $<$ 20\%) in Niger, where U/R stratified model reduces the bias from 3.4 to 1.2 per 1000 births.


Due to data sparsity, only cluster-level models produce reliable estimates at admin-2 level. We compare the yearly estimates of U5MR at admin-2 level from the model incorporating U/R stratification and those from the model without U/R stratification. Figure \ref{fig:u5mr-admin2-all-year} shows that U/R stratification lead to higher U5MR estimates in general, as we expected. The discrepancy between two sets of estimates suggests that adjusting for U/R stratification can make a difference. Extending the evidence from admin-1 to admin-2, we believe that U/R stratification helps to eliminate the underestimation of U5MR, although there is is no reasonable way to validate because survey estimates is not reliable at admin-2 level. Figure \ref{fig:u5mr-admin2-all-year} shows a map for the admin-2 level U5MR estimates from 2008 to 2012. We observe spatial structure in the U5MR estimates where neighboring areas share similar risk. We also see an overall temporal trend, in that U5MR gradually decrease over year. 


\subsection{Software Used for Computation}

The Bayesian hierarchical models for prevalence estimation utilizes integrated nested Laplace approximation (INLA) and we implement the models using R package INLA \citep{rue:etal:17}. The SUMMER package provides methods for spatial and spatio-temporal smoothing of demographic and health indicators using survey data \citep{li:etal:21}. We adopt the SUMMER package for the U5MR example and visualization. We implement BART and GBT using R package BART \citep{sparapani2021nonparametric} and 
bartMachine \citep{kapelner2013bartmachine}. Visualizations for Google maps are produced from the ggmap package in R\citep{kahle2013ggmap}.

\subsection{Model Details for Gradient boosted trees }

Gradient boosted trees (GBT) combine multiple regression trees based on certain loss function. For classification, we aim to minimize the error rate quantified as
$$\overline{\mbox{err}} = \frac{1}{N} \sum_{i=1}^N I(y_i \neq G(\bx_i)).$$
A classifier is sequentially applied to repeatedly modified versions of the data to produce a sequence of classifiers $G_m(\bx),m=1,2,\dots,M$, focusing on classifying cases that are poorly classified. Specifically, the data are modified at each step by applying weights $w_i$,  to the training observations $x_i,y_i$,$i=1,\dots,N$. Initially, the weights are set to $w_i=1/N$. Subsequently, at step $m$ ($m=2,\dots,N$), those observations that were misclassified by the classifier at the previous step, $G_{m-1}(\bx)$ are  upweighted, while those correctly classified are downweighted. 

The gradient boosted trees algorithm can be summarized as 

\noindent\hrulefill
\begin{enumerate}
\item Initialize weights $w_i=1/N$, $i=1,\dots,N$.
\item For $m=1,\dots,N$:
\begin{enumerate}
\item Fit a regression tree $G_m(\bx)$.
\item Compute
$$\mbox{err}_m = \frac{\sum_{i=1}^M w_i I( y_i \neq G_m(\bx_i)) }{\sum_{i=1}^M w_i }.$$
\item Compute $\alpha_m = \log[ (1-\mbox{err}_m)/\mbox{err}_m]$.
\item Set $$w_i \leftarrow w_i \exp [\alpha_m I(y_i \neq G_m(\bx_i))],$$ $i=1,\dots,N$.
\end{enumerate}
\item Output:
$$G(\bx) = \mbox{sign}\left( \sum_{m=1}^M \alpha_m G_m(\bx) \right).$$
\end{enumerate}
\vspace{-.03in}
\hrulefill

\subsection{Additional Modeling Choices}
\label{sec:method-extend}
\subsubsection{Covariates}

Borrowing information from covariates can boost the performance in area-level or cluster-level models \citep{rao:molina:15}. It is straightforward to incorporate area-level covariate in the modeling framework when we have mean covariate values at the same level as the target domain, such as admin-2. As an example, we can modify the prevalence model to
\begin{eqnarray}\label{eq:binomial}
Z_{c} | p_{c} &\sim& \mbox{Binomial}(n_{c},p_{c})
\nonumber \\
p_c  &=& \mbox{expit} \left(  \alpha + \bmx_{i[\bms_c]}\bbeta+ \gamma \times I( \bms_c \in \mbox{urban}) + S(\bolds_c)
+\epsilon_c \right) \label{eq:cov-prev-area}
\end{eqnarray}

Since the effect of covariate is identical for all individuals from the same area, there is no within-area variation in the risk beyong urban/rural stratification. Aggregation of urban/rural specific risk is the same as the procedure for models without covariates. 

Area-level covariate models are simple to implement, but are subject to ecological bias \citep{wakefield:08}. On the other hand, cluster-level covariate models are closer to the true underlying mechanism of action. The corresponding risk model can be modified to
\begin{eqnarray}\label{eq:binomial}
Z_{c} | p_{c} &\sim& \mbox{Binomial}(n_{c},p_{c})
\nonumber \\
p_c  &=& \mbox{expit} \left(  \alpha + \bmx(\bms_c)\bbeta+ \gamma \times I( \bms_c \in \mbox{urban}) + S(\bolds_c)
+\epsilon_c \right) \label{eq:cov-prev-cluster}
\end{eqnarray}
The aggregated prevalence for area $i$ is
\begin{equation}\label{eq:cov-prev-aggre} p_i = \int_{A_i} p(\bolds) \times N(\bolds)~d\bolds~ \approx ~ \sum_{l=1}^{M_i} p(\bolds_l) \times N(\bolds_l)
\end{equation}
where $N(\bolds)$ is the population density at $\bms$. We discretize the space by grid (pixels) indexed by $l = 1, . . . , M_i$ and use numerical integration to approximate the integral. 

In fact, aggregation for U/R specific risk without incorporating covariate is a special case of the above formulation -- when there are only two strata, the numerical integration becomes a linear combination of the two strata specific risks. \cite{wu:etal:21} noted that in general, for binary response model, covariate information is required for all individuals (locations) in the aggregation step because cluster-level probabilities undergo nonlinear transformation. Thus, no simplification may be made when incorporating additional covariates, especially continuous ones. Moreover, it may be challenging to obtain complete covariate information for the entire population, making cluster-level covariate models less appealing in LMICs setting.


\subsubsection{Continuous Model}

The models we implement are based on discrete or areal spatial effects. It is possible to adopt the model-based geostatistical approach to model spatial pattern in continuous space and/or time. A comprehensive introduction can be found in \cite{diggle:giorgi:19}. 

Here we briefly describe an application of continuous spatial modeling for a binary indicator. Instead treating $S(\bolds_c)$ as area-level spatial effects, we model it as a Mat\'ern covariance function \citep{stein:99}, parameterized by the marginal variance $\sigma_s^2$, the spatial range $\rho_s$, and 
the smoothness $\nu_s$. Various methods have been proposed to make the estimation process computationally feasible. For example, \cite{Lindgren:etal:11} pioneered the stochastic partial differential equations (SPDE) approach coupled with {\tt INLA} \citep{rue:etal:09,lindgren:rue:15}. 

If we include U/R as a covariate in the continuous spatial model, we can obtain strata specific risk maps. Since our U/R classification map is also at fine-scale (pixel level), we can combine then into a single risk surface. An additional aggregation step as described in Equation (\ref{eq:cov-prev-aggre}) will lead us to area-level prevalence estimate. However, we recommend modeling directly at the spatial level of interest and avoid aggregating from finer scales, as \cite{paige:etal:20} has shown that at target level, discrete spatial models perform as well as continuous-space models, but are far more straightforward to implement.

\subsection{Details for Prevalence Model (Space-only)}
\subsubsection{Area-level Fixed Effects}
\begin{enumerate}
    \item 
Without U/R adjustment: $$\mbox{logit}(p_{c})= \alpha_{i[\bms_c]} + \epsilon_c$$
    \item 
With U/R intercept: $$\mbox{logit}(p_{c})= \alpha_{i[\bms_c]}+\gamma \times I( \bms_c \in \mbox{urban}) + \epsilon_c$$
    \item 
Full U/R adjustment: $$ \mbox{logit}(p_{c})=\alpha_{i[\bms_c]} + \gamma_{i[\bms_c]} \times I( \bms_c \in \mbox{urban}) + \epsilon_c$$ 
\end{enumerate}

\subsubsection{Area-level IID Random Effects}

\begin{enumerate}
    \item 
Without U/R adjustment: $$\mbox{logit}(p_{c})= \alpha+ e_{i[\bms_c]} + \epsilon_c$$
    \item 
With U/R intercept: $$\mbox{logit}(p_{c})= \alpha +\gamma \times I( \bms_c \in \mbox{urban}) +  e_{i[\bms_c]}+\epsilon_c$$
    \item 
Full U/R adjustment:
\begin{eqnarray*}
\mbox{logit}(p_{c})  &=& \alpha +  e_{i[\bms_c]}^\text{rural} \times I( \bms_c \in \mbox{rural})+ \nonumber\\
 && (\gamma+ e_{i[\bms_c]}^\text{urban}) \times I( \bms_c \in \mbox{urban}) +\epsilon_c 
\end{eqnarray*}

\end{enumerate}

\subsubsection{Area-level Spatial Random Effects}

\begin{enumerate}
    \item 
Without U/R adjustment: $$\mbox{logit}(p_{c}) =   \alpha + S(\bolds_c) +\epsilon_c $$

\item With U/R intercept:
$$\mbox{logit}(p_{c}) =   \alpha + \gamma \times I( \bms_c \in \mbox{urban}) + S(\bolds_c) +\epsilon_c $$
\item Full U/R adjustment:
\begin{eqnarray}
\mbox{logit}(p_{c})  &=& \mbox{expit} \Bigl( \alpha +  S^\text{rural}(\bolds_c)\times I( \bms_c \in \mbox{rural})
)+\nonumber\\
 &&  (\gamma+S^\text{urban}(\bolds_c)) \times I( \bms_c \in \mbox{urban}) +\epsilon_c 
\Bigr)
\end{eqnarray}

\end{enumerate}

\begin{table} 
\caption{\label{table:HIV-uncertainty} Average 95\% CI widths for HIV prevalence of women of age 15-49 in Malawi, from various models. Numbers are infections per 100 women. }
\centering
\begin{tabular}{llrrrr}
  \hline
 Model type & Stratification & \makecell[l]{CI width from \\ fixed fractions} & \makecell[l]{CI width from \\ sampled fractions} & \makecell[l]{CI width \\ difference}\\ 
  \hline
 Fixed effects & U/R+Area & 7.05 & 7.14 & 0.08 \\ 
 Fixed effects & U/R*Area & 7.37 & 7.48 & 0.11 \\ 
 Area IID & U/R+Area  & 6.83 & 7.00 & 0.17 \\ 
 Area IID & U/R*Area & 6.87 & 6.96 & 0.09  \\ 
 BYM2 & U/R+Area & 5.98 & 6.03 & 0.06 \\ 
 BYM2 & U/R*Area & 6.14 & 6.16 & 0.01 \\ 
   \hline
\end{tabular}
\end{table}

\subsection{Quantify Uncertainty}

Building the entire pipeline under a Bayesian framework enables us to properly evaluate uncertainties. In this section, we measure the uncertainties of estimated urban fractions from the classification model, and investigate how the variation is propagated to the estimates of health indicators. 

As an example, we estimate the posterior of urban fraction in the 27 admin-2 regions in Malawi (Likoma excluded). Just for illustration purpose, we use BART-MCMC as the classification model and total population in 2008 as the target population. To obtain samples of urban fractions, we need samples for the classification surface. The exact benchmark method described in Equation (\ref{eq:exact-benchmark}) cannot guarantee that the central of the posterior for urban fractions agrees with the known truth from the census. Thus, we conduct a grid search to find the cutoff $\tau_i$ for each admin-2 area. 

\begin{figure} [H]
    \centering
    \includegraphics[width=0.95\linewidth,trim={0 0cm 0 0.8cm},clip]{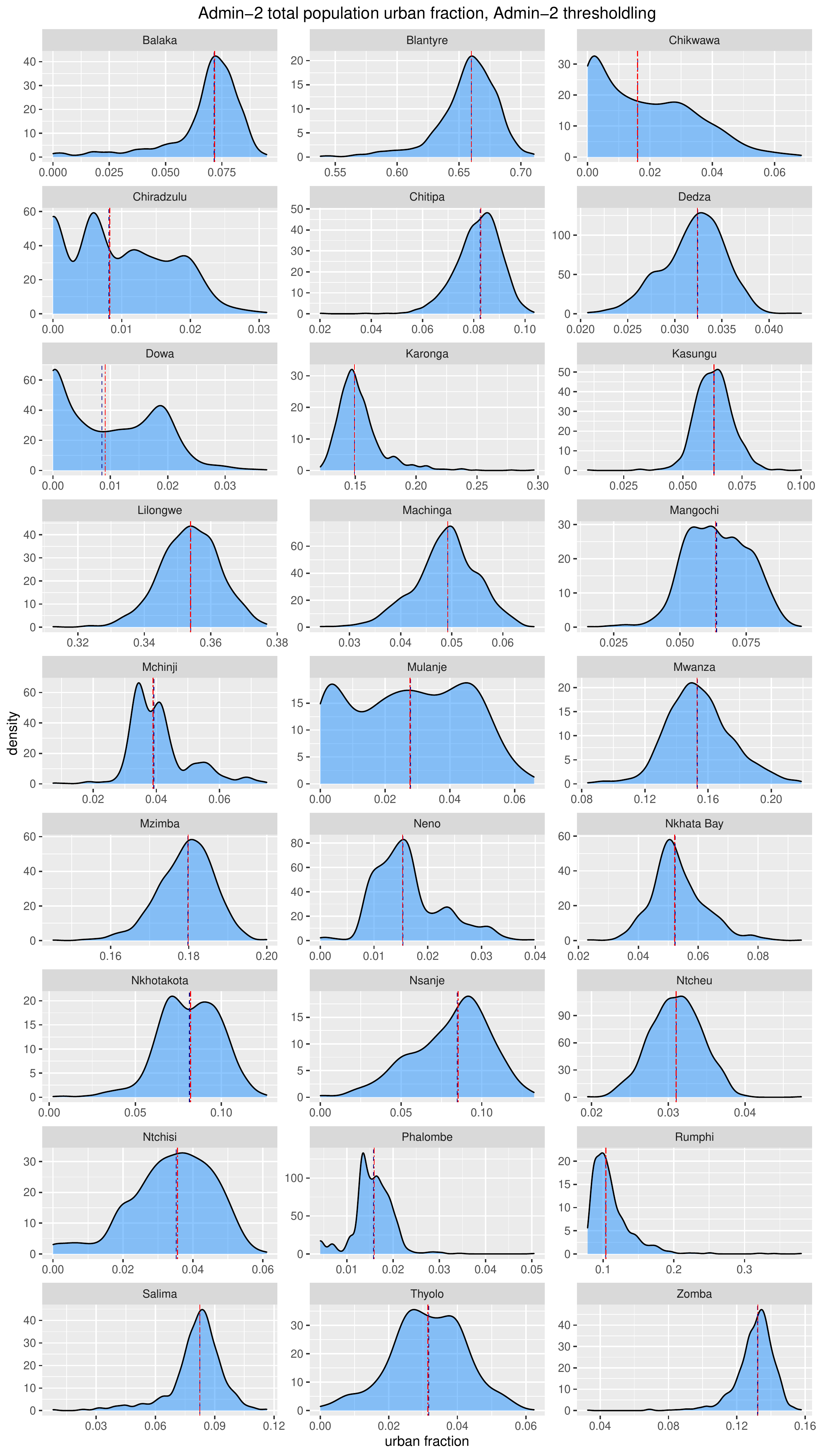}  
    \caption{Density plots representing the posteriors for urban fractions for admin-2 regions in Malawi}
    \label{fig:uncertainty-total-pop}
\end{figure}

Figure \ref{fig:uncertainty-total-pop} shows the posterior density of urban fractions for the total population in each admin-2 region in 2008. The posterior median and known fractions, represented by blue and red vertical lines, match well. The widths of 95\% CIs for all admin-2 regions are smaller than 0.10, with majority smaller than 0.05. The tight CIs show that our estimates have high precision such that U/R stratification is not expected to bring much extra variation to the prevalence estimates.

To quantify the increase of uncertainties introduced by urban fractions, we revisit the HIV prevalence example in Section \ref{sec:HIV-example}. Earlier we used point estimates of urban fractions (fixed) for aggregation, but here we consider samples of urban fractions in the aggregation procedure. Equation (\ref{eq:fixed-r}) corresponds to the aggregation method for the third column in Table \ref{table:HIV-uncertainty}, while Equation (\ref{eq:sampled-r}) corresponds to the fourth column. 

\begin{eqnarray}
p^{(m)}_i &=&  p^{(m)}_{i,U}\times q_{i} + p^{(m)}_{i,R}\times (1-q_{i})
\label{eq:fixed-r}
\\
p^{(m)}_i &=&  p^{(m)}_{i,U}\times q^{(m)}_{i} + p^{(m)}_{i,R}\times (1-q^{(m)}_{i})
\label{eq:sampled-r}
\end{eqnarray}

We average the widths of 95\% CIs across 27 admin-2 areas in Malawi (Likoma excluded) and display them in Table \ref{table:HIV-uncertainty}. The uncertainties for HIV prevalence increased slightly when the uncertainties from the classification model in incorporated. Across all models, the increase of CI width is less than 0.2, which corresponds to less than 0.2 infections per 100 women. 

Based on the above evidence, we conclude that the extra variation brought by estimating urban fraction is negligible. In general, we recommend using point estimates for urban fractions in the aggregation. Extra caution might be needed for very fine spatial resolution such as admin-3 level (or certain  particularly small admin region), as the uncertainties for the urban fractions will be higher. We also note that there are additional source of uncertainty that we did not measure, in particular from the population density file and covariates (which we treat as fixed).

\subsection{Likoma Exclusion Justification}

Likoma District is the smallest and least populous admin-2 region in Malawi. It is primarily constituted by two islands, Likoma and Chizumulu, in Lake Malawi. Due to its special geographical location and small area, the shapefiles and population density surface do not agree well in terms of the boundary of the islands. In particular, the islands are too small to contain more than a few 1km$\times$1km pixels. The urban proportion for the total population in Likoma region is about 10$\%$, according to the census. However, based on the population density surface, more than 50$\%$ of the population is concentrated within one single pixel. All (reasonable) classification models will determine $\bms_g$ as more likely to be urban than any other pixel in Likoma region. Thus, we cannot properly recover the 10$\%$ urban fraction from the census.

In the Likoma region, the population density surface might wrongly allocated population into pixel $\bms_g$. As, we rely heavily on the precision of the population density surface, it is infeasible to solve the issue from a statistical perspective. It might also be the case that the true urban boundary lies entirely within pixel $\bms_g$, but there is no way to verify. We emphasize that Likoma is an edge case. Pixels of resolution 1km $\times$ 1km are fine enough for almost all circumstances. 

Based on the above reasons, we do not apply our estimation method for Likoma region. Thus we exclude it in comparing precision of estimated admin-2 urban fractions across methods, and in quantifying uncertainties of estimated urban fractions. For estimating admin-2 level HIV prevalence, we use the known urban fraction of total population at 2008 (the year of census) as the best approximate for the urban fraction of women of age 15-49 at 2015. 

\newpage

\subsection{Additional Figures}

\begin{figure}[H]
    \centering
    \includegraphics[width=1\linewidth,trim={0 0cm 0cm 0},clip]{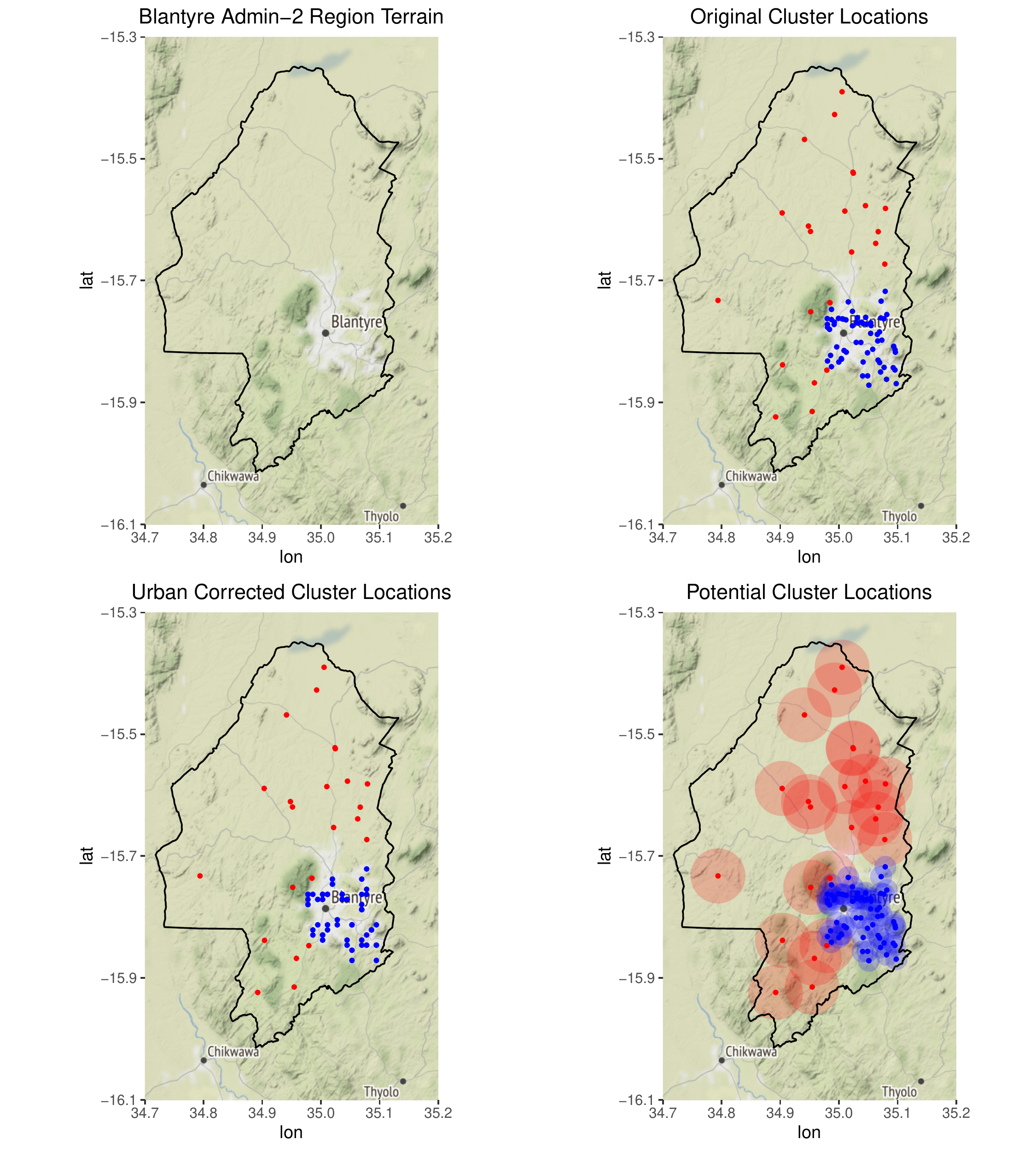}
      \caption{Clusters in Blantyre. Urban and rural clusters are labeled blue and red respectively. }
    \label{fig:cluster_Blantyre}
\end{figure}

Figure \ref{fig:cluster_Blantyre}(b) and (c) displays the cluster locations before and after the correction respectively. We emphasize that only the urban clusters are shifted and the rural clusters remain unchanged. 
 
\newpage

\begin{figure}  [H]
    \centering
    \includegraphics[width=0.95\linewidth,trim={0 0cm 0 0cm},clip]{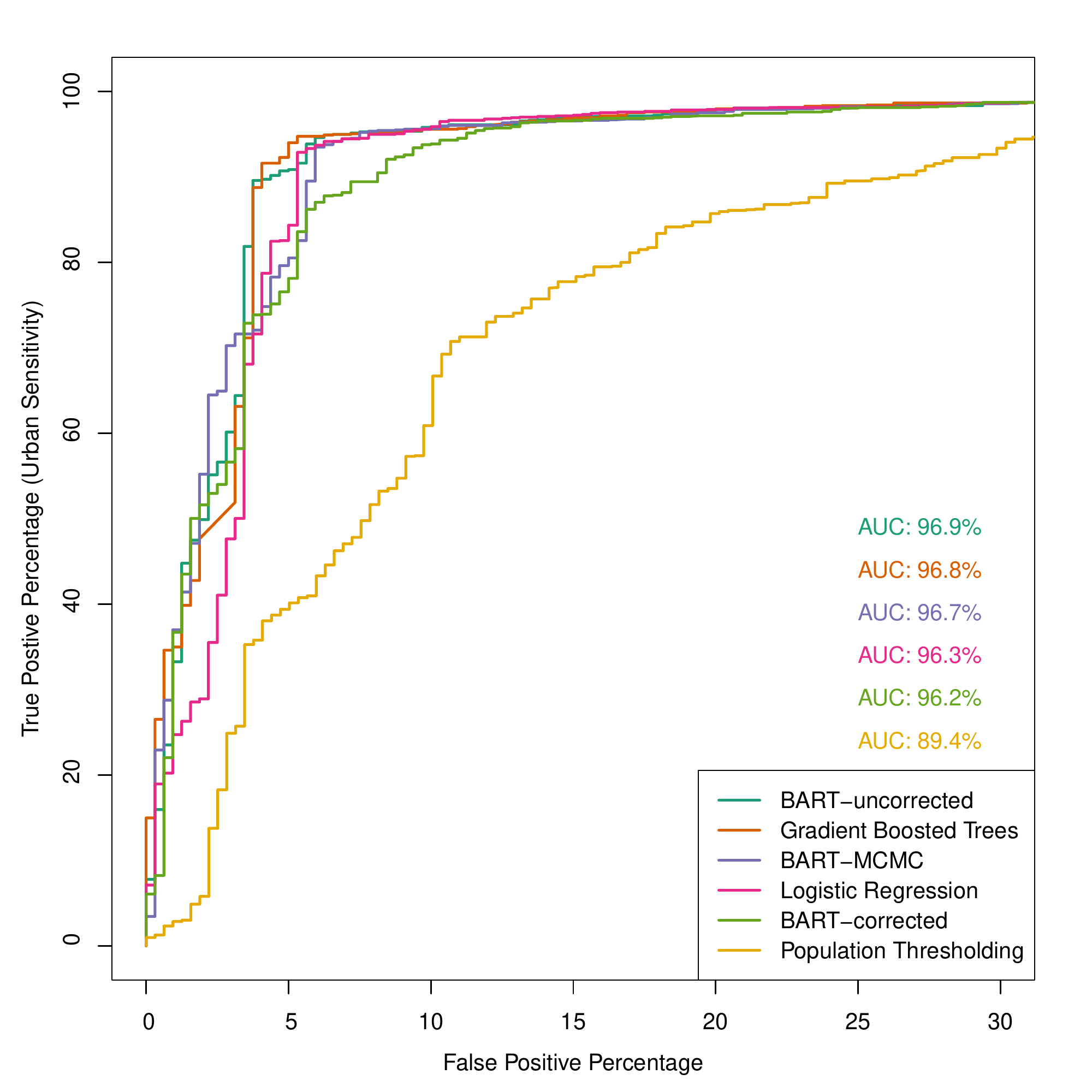}  
    \caption{Zoomed ROC curve for 5-fold CV across methods}
    \label{fig:ROC-curves-zoomed}
\end{figure}

Figure \ref{fig:ROC-curves-zoomed} shows the zoomed cross validation ROC curves. Models except for the naive population thresholding method share similar performance.

\begin{figure}  [H]
    \centering
    \includegraphics[width=0.5\linewidth,trim={0 0cm 0 0cm},clip]{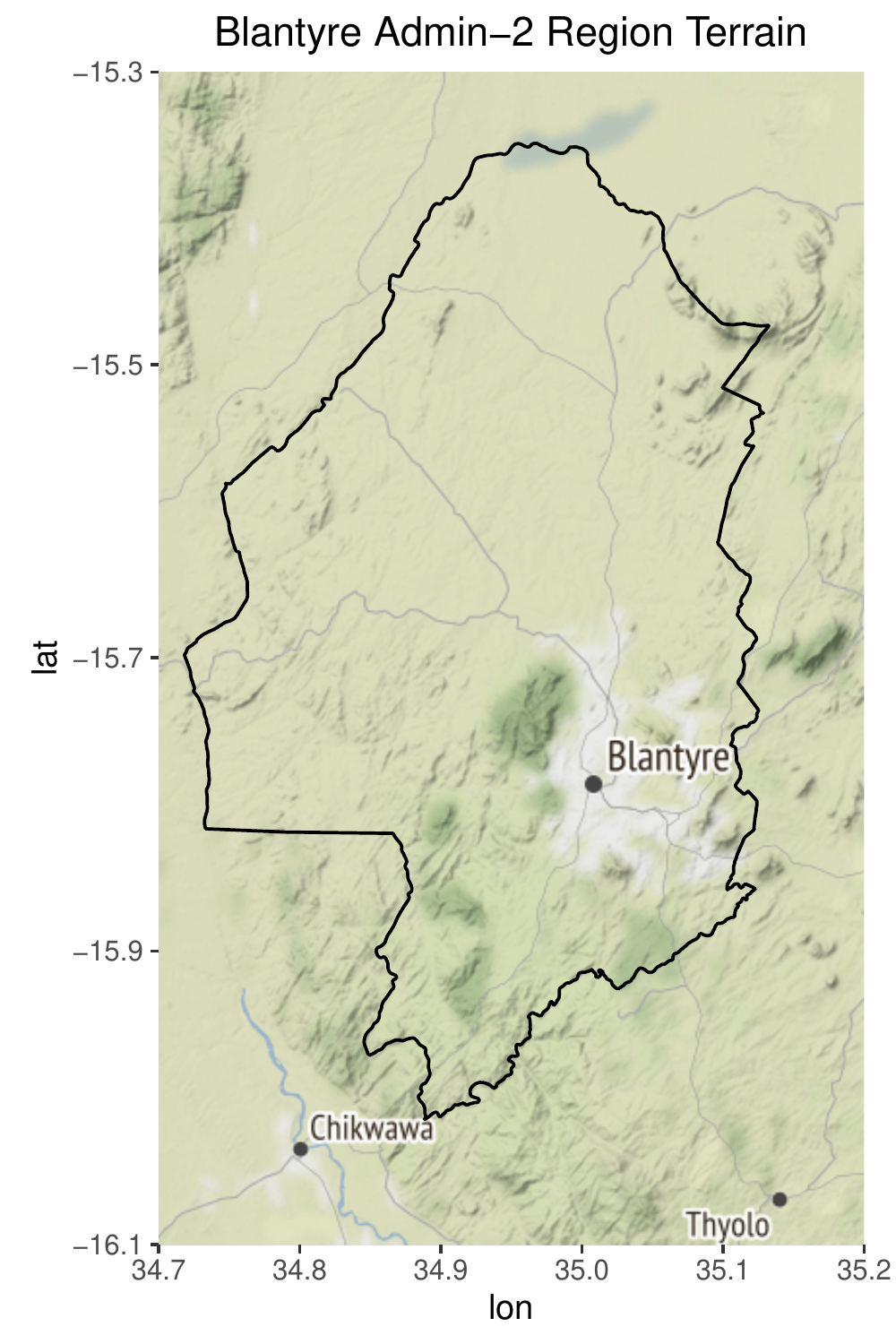}
     \caption{Blantyre terrain map, both urban and rural included}
    \label{fig:Blantyre-map}
\end{figure}

\newpage
Figure \ref{fig:Blantyre-map} shows a terrain map for the admin-2 region Blantyre in Malawi. The light colored region can serve as an approximate for Blantyre city (urban part of Blantyre).

\begin{figure}  [H]
    \centering
    \includegraphics[width=0.95\linewidth,trim={0 0cm 0 0cm},clip]{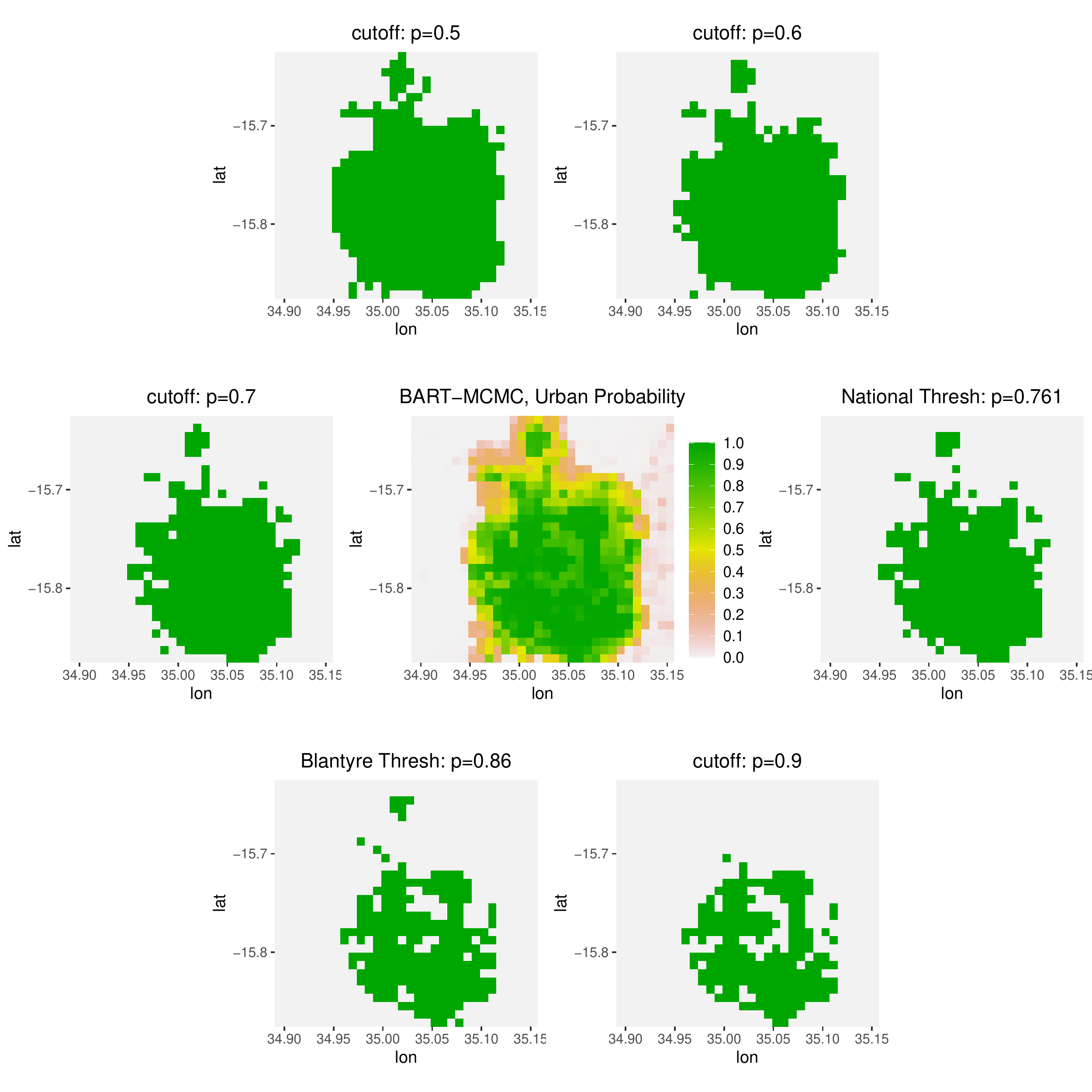}
         \caption{Blantyre predicted probability map, along with maps for urban classifications using various cutoffs.}
    \label{fig:Blantyre-jitter-prob}
\end{figure}

\newpage
Figure \ref{fig:Blantyre-jitter-prob} visualizes the estimated urban regions using different probability cutoffs, based on BART-MCMC model.

\newpage

\begin{figure} [H]
    \centering
    \makebox{
    \includegraphics[width=0.95\linewidth,trim={0 0cm 0 0cm},clip]{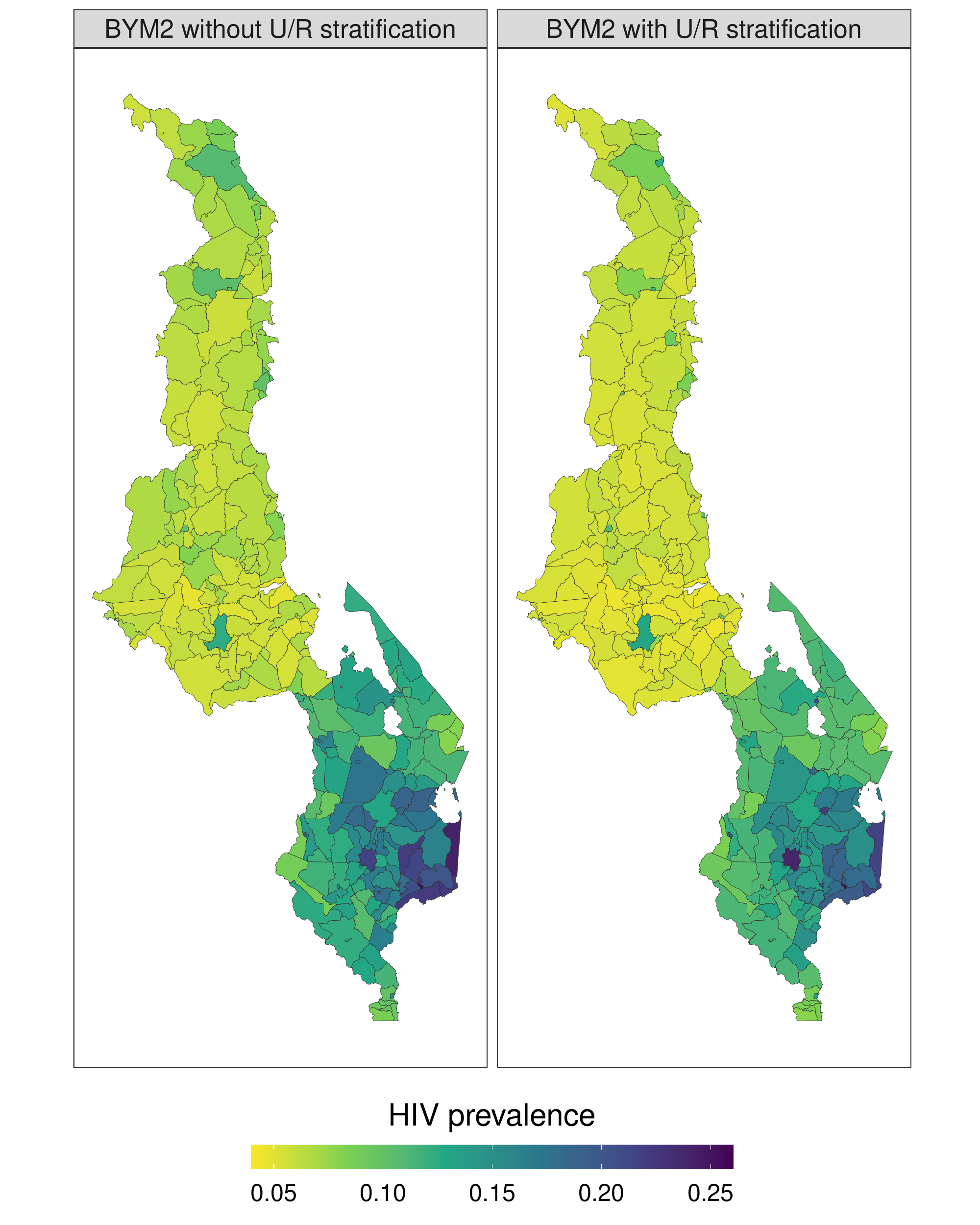}
    }
    \caption{Maps for HIV prevalence estimates at admin-3 level in Malawi from BYM2 model with and without U/R stratification.}
        \label{fig:HIV-admin3-map}
\end{figure}

Figure \ref{fig:HIV-admin3-map} suggests that stratified model yield lower HIV prevalence estimates in most regions, indicated by the lighter color. The difference is expected because oversampling of urban samples is present in most areas in Malawi.

\begin{figure} [H]
    \centering
    \makebox{
    \includegraphics[width=0.95\linewidth,trim={0 0cm 0 0cm},clip]{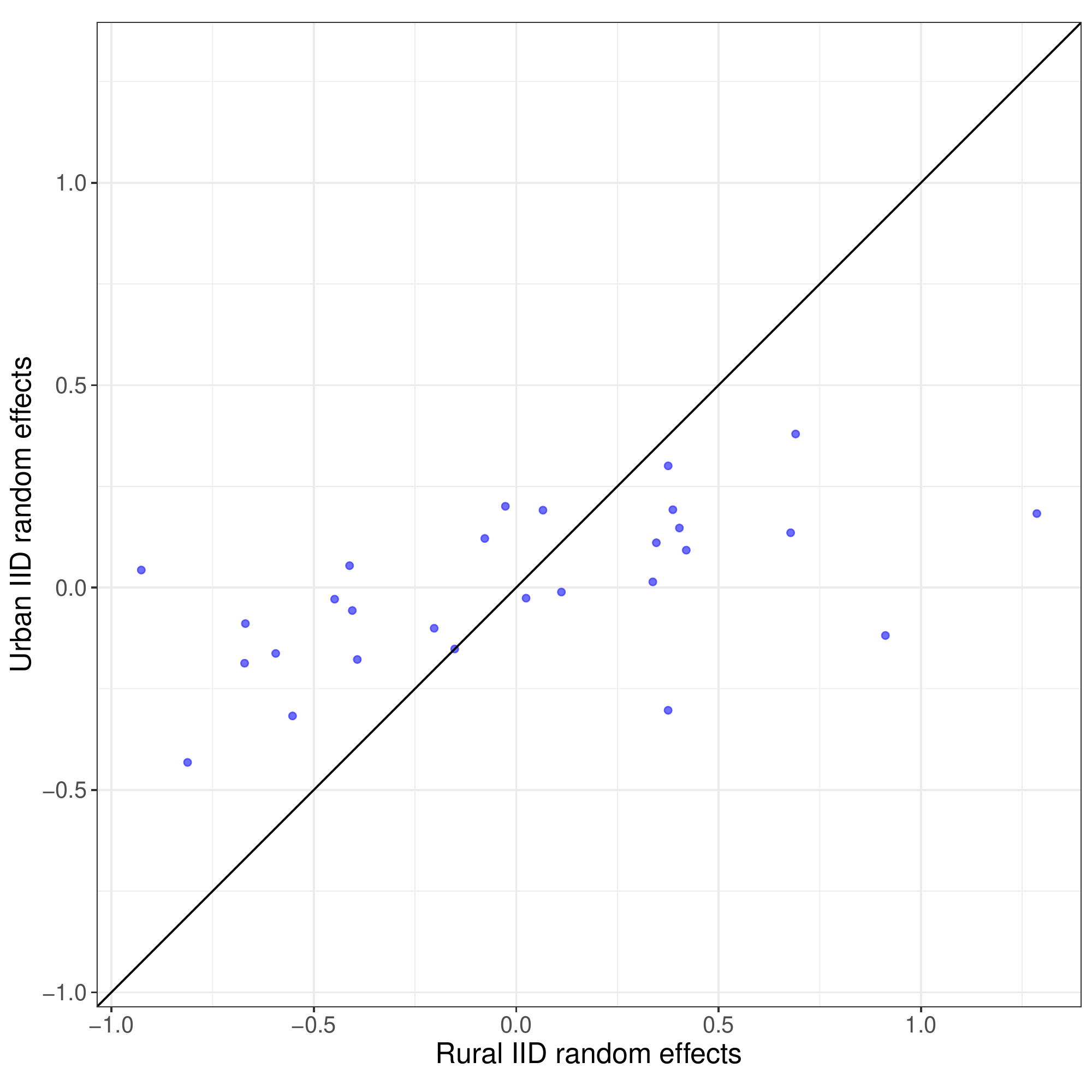}
    }
    \caption{Comparison of urban/rural specific area-level random effects from area-level random effects model with full U/R stratification.}
        \label{fig:IID-UR-RE-scatter}
\end{figure}

\begin{figure} [H]
    \centering
    \makebox{
    \includegraphics[width=0.95\linewidth,trim={0 0cm 0 0cm},clip]{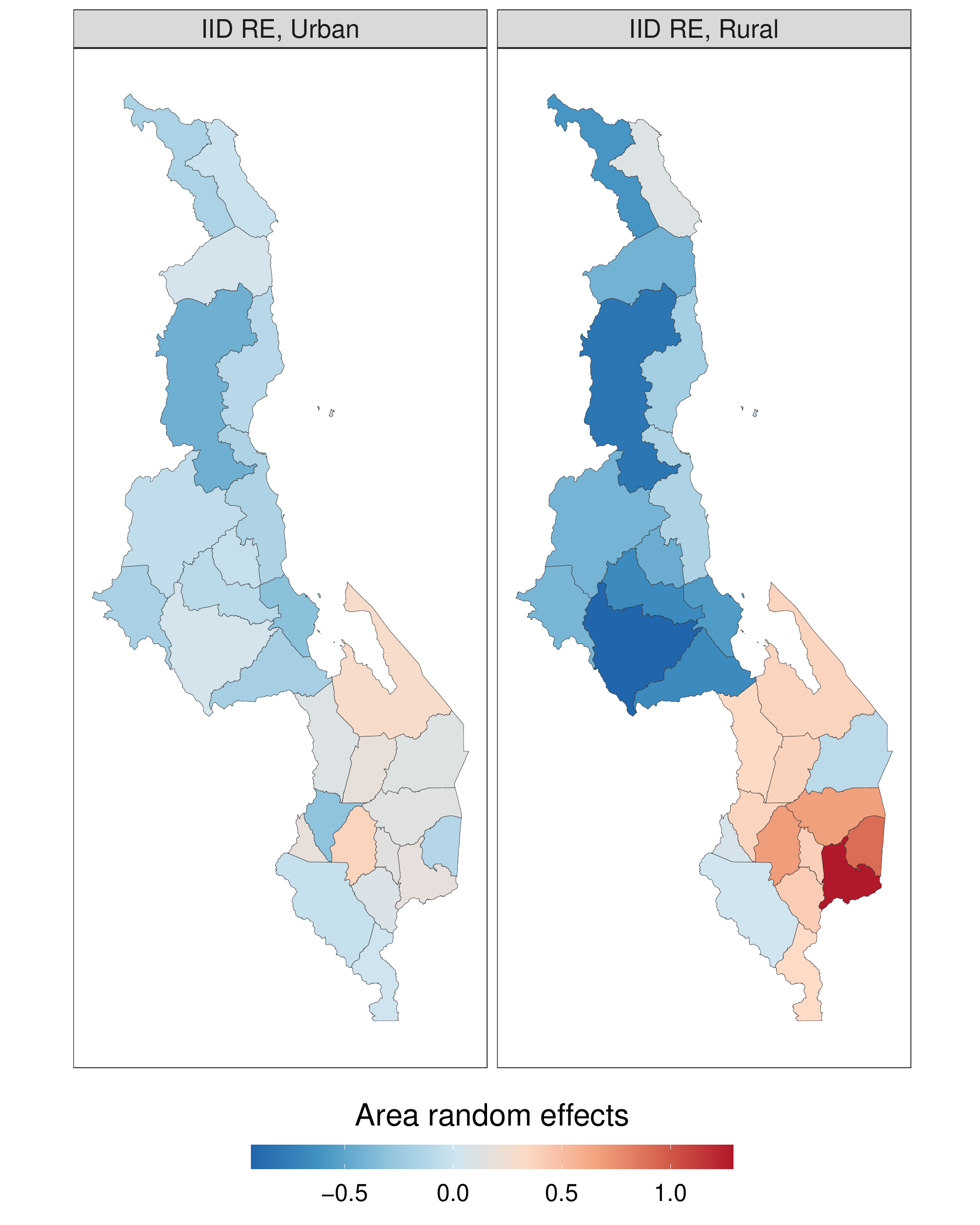}
    }
    \caption{Maps for urban/rural specific area-level random effects from area-level random effects model with full U/R stratification.}
        \label{fig:IID-UR-RE}
\end{figure}

\newpage
Figure \ref{fig:IID-UR-RE-scatter} and \ref{fig:IID-UR-RE} indicate that the magnitude of IID random effects is higher for rural areas compared with their urban counterparts. We expected such results because rural samples are richer so that there would be less smoothing. The overall spatial patterns for rural and urban areas are similar -- HIV risks are higher in southern regions and lower in nortern regions.

\begin{figure} [H]
    \centering
    \makebox{
    \includegraphics[width=0.95\linewidth,trim={0 0cm 0 0cm},clip]{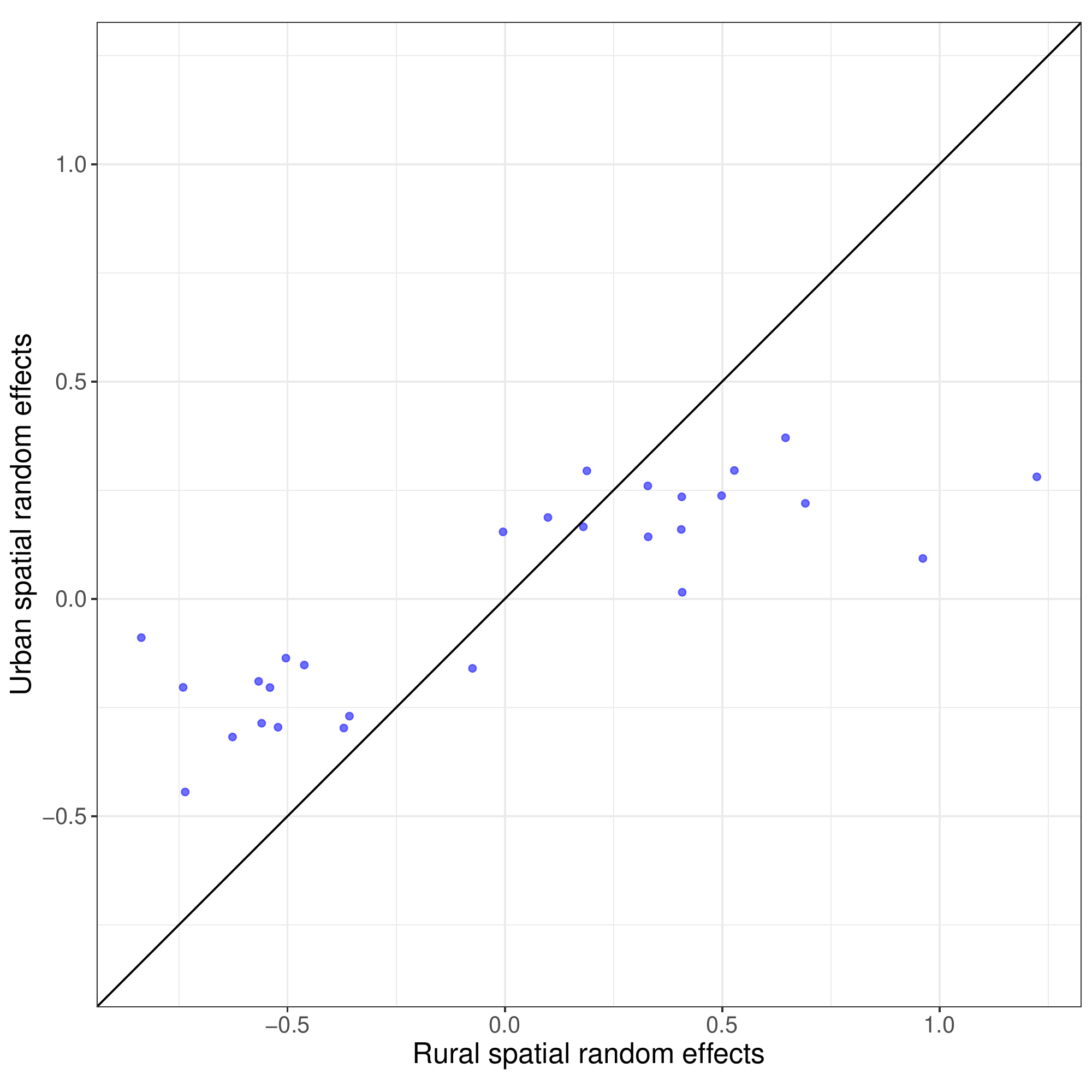}
    }
    \caption{Maps for urban/rural specific area-level random effects from BYM2 model with full U/R stratification.}
        \label{fig:BYM2-UR-RE-scatter}
\end{figure}

\begin{figure} [H]
    \centering
    \makebox{
    \includegraphics[width=0.95\linewidth,trim={0 0cm 0 0cm},clip]{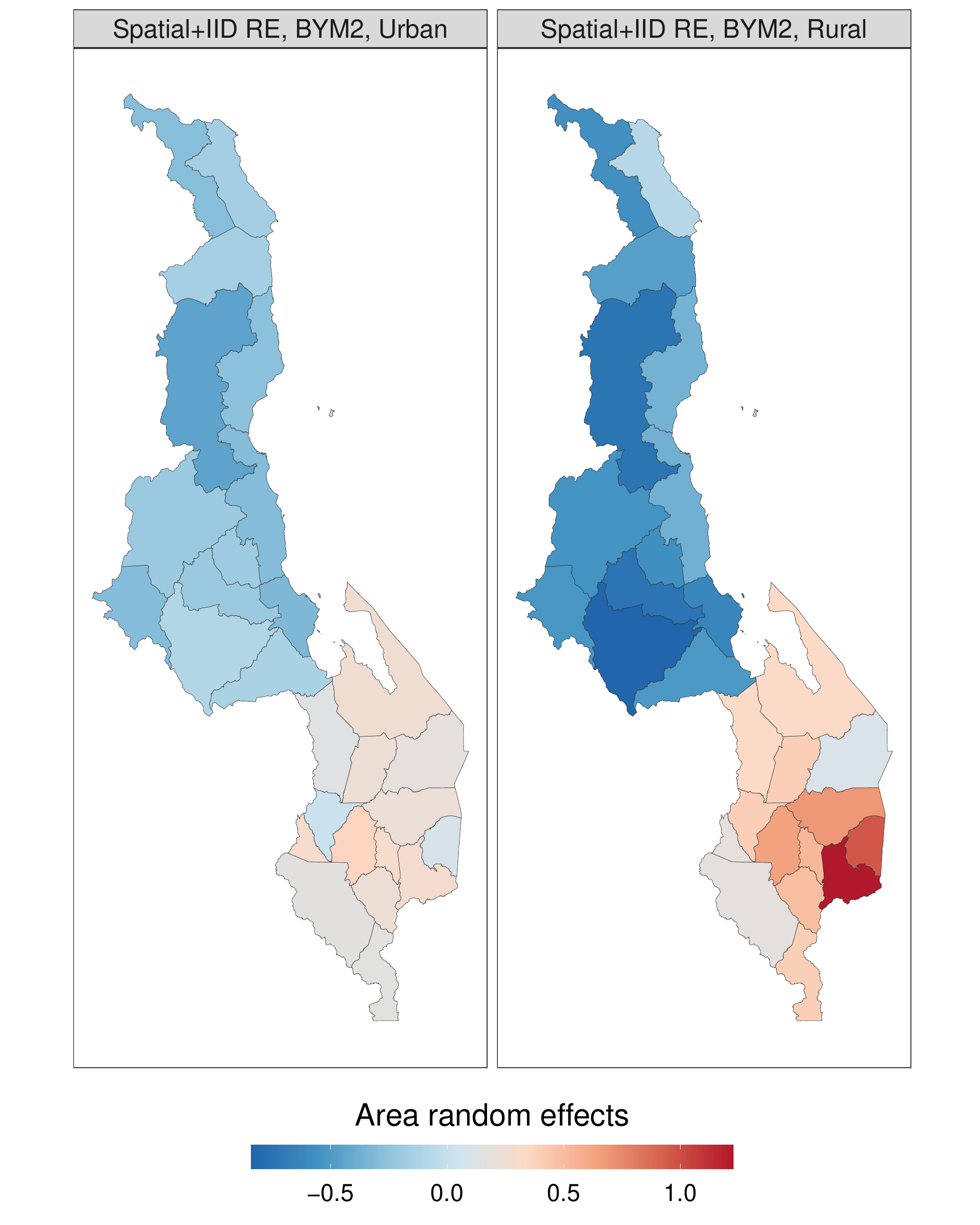}
    }
    \caption{Maps for urban/rural specific area-level random effects from BYM2 model with full U/R stratification.}
        \label{fig:BYM2-UR-RE}
\end{figure}

\newpage
Figure \ref{fig:BYM2-UR-RE-scatter} and \ref{fig:BYM2-UR-RE} plot the random effects for BYM2 models. Here the random effects constitute of both the IID component and the spatial component. We observe results similar to those from the area-level random effects model.

\end{document}